\documentclass[
	nofootinbib,
         amsmath,
         amssymb,
         superscriptaddress,
         aps,
         prb,
         longbibliography,
         floatfix,
         twocolumn
]{revtex4-2}
\usepackage[english]{babel}
\usepackage[utf8]{inputenc} 
\usepackage{dsfont}
\usepackage{SIunits}
\usepackage[caption=false,label font=bf,labelformat=simple]{subfig}
\usepackage{graphicx}     
\usepackage{color}
\usepackage[normalem]{ulem}
\usepackage{soul}
\usepackage{cancel}
\usepackage{hyperref}

\usepackage[normalem]{ulem}

\begin{document}
\title{Probing anyon statistics on a single-edge loop in the fractional quantum Hall regime}
\author{Flavio Ronetti}
\affiliation{Aix Marseille Univ, Universit\'e de Toulon, CNRS, CPT, Marseille, France}
\author{Noé Demazure}
\affiliation{Aix Marseille Univ, Universit\'e de Toulon, CNRS, CPT, Marseille, France}
\author{Jérôme Rech}
\affiliation{Aix Marseille Univ, Universit\'e de Toulon, CNRS, CPT, Marseille, France}
\author{Thibaut Jonckheere}
\affiliation{Aix Marseille Univ, Universit\'e de Toulon, CNRS, CPT, Marseille, France}
\author{Benoît Grémaud}
\affiliation{Aix Marseille Univ, Universit\'e de Toulon, CNRS, CPT, Marseille, France}
\author{Laurent Raymond}
\affiliation{Aix Marseille Univ, Universit\'e de Toulon, CNRS, CPT, Marseille, France}
\author{Masayuki Hashisaka}
\affiliation{Institute for Solid State Physics, University of Tokyo, 5-1-5 Kashiwanoha, Kashiwa, Japan}
\author{Takeo Kato}
\affiliation{Institute for Solid State Physics, University of Tokyo, 5-1-5 Kashiwanoha, Kashiwa, Japan}
\author{Thierry Martin}
\affiliation{Aix Marseille Univ, Universit\'e de Toulon, CNRS, CPT, Marseille, France}

\newcommand{\lr}[1]{\textcolor{green}{#1}}
\definecolor{Jerome}{rgb}{0.8, 0.0, 0.3}
\newcommand{\jr}[1]{\textcolor{Jerome}{#1}}

\begin{abstract}
We propose a setup to directly measure the anyonic statistical angle on a single edge of a fractional quantum Hall system, without requiring independent knowledge of non-universal parameters. We consider a Laughlin edge state bent into a closed loop geometry, where tunneling processes are controllably induced between the endpoints of the loop. To illustrate the underlying physical mechanism, we compute the time-dependent current generated by the injection of multiple anyons, and show that its behavior exhibits distinctive features governed by the anyonic statistical angle. The measured current reflects quantum interference effects due to the time-resolved braiding of anyons at the junction. To establish experimental relevance, we introduce a protocol where anyons are probabilistically injected upstream of the loop via a quantum point contact (QPC) source. Unlike in Fabry–Perot interferometers, where phase jumps occur spontaneously due to stochastic quasi-particle motion, here the phase jumps are deliberately induced by source injections. These events imprint measurable signatures in the cross-correlation noise, enabling a controlled {statistical} analysis of the braiding phase. We further show that, by varying the magnetic field while remaining within the same fractional quantum Hall plateau, the statistical angle can be extracted without relying on the knowledge of other non-universal system parameters. Our results provide a minimal and accessible platform for probing anyonic statistics using a single chiral edge.
\end{abstract}
\let\endtitlepage\relax
\maketitle
\section{Introduction}
Quantum statistics classify identical particles based on the behavior of their wavefunctions under particle exchange. In three-dimensional space, only two possibilities exist: the wavefunction is symmetric for bosons or antisymmetric for fermions, defining the two fundamental classes of quantum particles. However, the topology of two-dimensional space allows for more exotic possibilities—exchanging identical particles can cause the wavefunction to acquire an arbitrary phase, giving rise to particles with generalized statistics known as anyons.~\cite{Leinaas77,Wilczek82,Arovas84}. 

One of the most prominent physical realizations of generalized exchange statistics is the fractional quantum Hall effect (FQHE)\cite{Tsui82,Laughlin83}, where electrons confined to two dimensions under strong magnetic fields form highly correlated topological phases. In these systems, the elementary excitations are anyons, which not only exhibit fractional statistics but also carry a fraction of the electron’s charge. This latter property is a hallmark of the FQHE and has been robustly confirmed over the past three decades through shot noise measurements\cite{Kane94,Saminadayar97,dePicciotto97,Martin05,Biswas22} and complementary studies employing alternative techniques~\cite{Goldman95,JensMartin04,Kapfer19,Bisognin19}. In contrast, the detection of fractional statistics has been achieved only recently, owing to the greater experimental complexity required by the underlying theoretical proposals~\cite{Chamon97,Safi01,Vishveshwara03,Law06,Bishara08,Halperin11,Campagnano12,Rosenow12,Levkivskyi12,Campagnano13,Lee19,Carrega21}.

These experimental breakthroughs provided compelling evidence for the fractional statistics of anyons in FQHE systems, particularly at filling factor $\nu=1/3$, through two main apparatus, the Fabry-Perot interferometer~\cite{Nakamura20} and the anyon collider~\cite{Bartolomei20}. 

The Fabry–Pérot geometry consists of two tunneling constrictions, known as quantum point contacts (QPCs), formed by negatively biased gates~\cite{Chamon97,Halperin11,McClure12,Carrega21}. At these junctions, quasi-particles can tunnel between edge states. The presence of two QPCs creates a closed-loop path, allowing for quantum interference between different tunneling trajectories. This interference is sensitive to the magnetic flux threading the loop, leading to observable changes in transport properties. Furthermore, if bulk quasi-particles become trapped within the interferometer area, they contribute additional phase shifts to the interference pattern~\cite{Camino05,Roosli20}. Specifically, when quasi-particles are backscattered at the QPCs, their paths encircle those localized in the loop, effectively realizing the anyon braiding thought experiment in a natural setting. Varying the number of enclosed quasi-particles results in measurable shifts in the interference phase—a clear signature of anyonic statistics. Experimental evidence for such fractional statistics has been reported across various fractional quantum Hall states and material platforms~\cite{Nakamura19,Nakamura20,Ronen21,Nakamura23,Kim24,Kim24b,Werkmeister24,Samuelson24}.

In contrast, the collider geometry is designed to investigate time-domain anyon braiding by triggering the injection of quasi-particles from a source QPC toward a central QPC~\cite{Rosenow16,Han16,Lee22,Morel22,Mora22,Schiller23,Jonckheere23}. When the source QPC is biased with a constant voltage and tuned to the weak backscattering regime, it emits anyons as rare, Poisson-distributed events. These anyons then propagate to the central QPC—also in the weak backscattering regime—where quasi-particle–quasi-hole pairs can be generated spontaneously. A key feature of this setup is the possibility that an incoming anyon reaches the central QPC either before or after the creation of such a pair, leading to quantum interference between two processes with distinct statistical phases. The resulting phase difference depends on the statistical angle $\pi \lambda$, a topologically protected quantity insensitive to sample-specific details. In contrast, the probability of spontaneous pair creation follows a temperature-dependent power law, governed by the scaling dimension $\delta$. Unlike the statistical angle, $\delta$ is not topologically protected and can be renormalized by interactions or other sample-dependent effects~\cite{Braggio12}.

Despite considerable theoretical efforts to develop alternative methods for extracting the scaling dimension of anyons in the fractional quantum Hall effect (FQHE)~\cite{Snizhko15,Rech20,Ebisu22,Schiller22,Bertin23,Iyer23,Acciai24}, experiments have yet to yield a definitive measurement~\cite{Veillon24,Schiller24,Ruelle24}. The scaling dimension manifests in the power-law relationship between current and voltage predicted by chiral Luttinger liquid theory. While some recent experiments have successfully extracted the theoretically predicted scaling dimension from this power-law behavior, many others conducted under similar conditions have failed to replicate these results. This disparity suggests that the measurement of the scaling dimension is complicated by non-universal, sample-dependent factors.

In this paper, we introduce a novel experimental setup designed to measure the statistical angle $\pi \lambda$ of topological anyons emitted by a source quantum point contact (QPC) without requiring an independent extraction of the scaling dimension $\delta$. Moreover, by taking inspiration from the Fabry-Perot setup, we show that a protocol can be engineered to extract the statistical angle by changing the magnetic flux without exiting the corresponding fractional quantum Hall (FQH) plateau. Our proposed system consists of a single edge of a FQH liquid at filling factor $\nu=1/\left({2n+1}\right)$, with $n \in \mathbb{N}$, incorporating an $\Omega$-shaped junction (see Fig.~\ref{fig:scheme}). This junction is engineered by geometrically defining a localized droplet within the FQH fluid. The proposed configuration draws inspiration from previous experiments conducted at integer filling factors in the context of mesoscopic capacitors~\cite{Gabelli07,Parmentier12}.

To explore the fundamental role of anyonic statistics in this setup, we compute the time-dependent charge current response to isolated delta-like anyon pulses injected into the system. Our analysis demonstrates that the temporal evolution of the charge current is governed primarily by the exchange statistics of anyons, while other non-universal parameters—such as temperature and the scaling dimension $\delta$—merely act to renormalize the current prefactor without affecting its qualitative features. The main braiding mechanisms responsible for the generation of the time-dependent current are elucidated. In addition, we focus on the case of the injection of two anyons separated by a fixed time delay and we show how the statistics and the relative positions of these anyons influence their time dynamics.

After having presented the main physical mechanism occurring at the loop-junction, we extend our study to a scenario closer to the collider geometry where a source QPC, positioned upstream of the $\Omega$-junction, emits a dilute stream of anyons when biased by a constant voltage $V_{DC}$. In this setup, we compute the finite-frequency current cross-correlations at two distinct output terminals of the system, placed downstream with respect to the source QPC. We show that these correlations provide a direct signature of anyon braiding, offering a robust method to extract the statistical angle $\pi \lambda$ without requiring prior knowledge of the scaling dimension. The latter can be obtained by measuring two zeros of the noise as a function of the bias $V_{DC}$ or as a function of the magnetic flux $\Phi$ threading the loop. This feature aligns with recent experimental protocols proposed to probe anyonic statistics in FQHE systems~\cite{Ruelle24}. The ability to isolate and measure the statistical phase independently represents a significant step toward the experimental validation of anyonic braiding in topological phases of matter. Moreover, our approach could open new avenues for the controlled manipulation of anyons in FQHE systems, contributing to the realization of topologically protected operations for quantum information processing.

The paper is organized as follows. In Section \ref{sec:Model} we introduce the Hamiltonian model describing our setup. Then in Sec. \ref{sec:CorrelationFunctions}, we present the correlation functions of the boson modes. This formalism is used to compute the time-dependent charge current in Section \ref{sec:Current}. The case of a dilute stream of anyons is considered in Section \ref{sec:Collider}, where we compute the finite-frequency noise that allows us to extract the parameter $\lambda$. Finally, in Section \ref{sec:Conclusions} we draw our conclusions. In the
following, units where $\hbar=1$ and $k_B=1$ are employed. Technical details of the calculations are provided in several Appendices.

\section{Model}
\label{sec:Model}
We focus on the edge of a quantum Hall bar operating in the fractional regime, characterized by a filling factor  
\begin{equation}
    \nu = \frac{1}{2n+1}, \quad n \in \mathbb{N}.
\end{equation}  
This system belongs to the Laughlin sequence, meaning that a single chiral edge channel is present, propagating along the boundary of the sample. The fundamental excitations of this edge state are fractional quasi-particles, which carry an elementary charge
\begin{equation}
    -e^* =- e \nu,
\end{equation}  
where $-e$ is the electron charge, and obey anyonic exchange statistics with a statistical angle given by  
\begin{equation}
    \pi \lambda = \pi \nu.
\end{equation}  

The low-energy properties of the chiral edge state can be described within the framework of conformal field theory (CFT), using a bosonized representation~\cite{Kane94,Wen95}. The Hamiltonian governing the dynamics of the single edge channel takes the form~\cite{Wen95}  
\begin{equation}
    \label{eq:HamEdge}
    H_0 = \frac{u}{4\pi} \int dx~\left[\partial_x\phi(x)\right]^2,
\end{equation}  
where $u$ denotes the velocity of the chiral bosonic mode, and $\phi(x,t)$ represents the chiral boson field, which satisfies the relation  
\begin{equation}
    \phi(x,t) = \phi(x-ut,0) = \phi(0,t-x/u).
\end{equation}

The chiral nature of the field is a direct consequence of the topology of the system and the presence of a strong perpendicular magnetic field. Moreover, the bosonic field obeys the fundamental commutation relation  
\begin{equation}
\left[\phi(x,t),\phi(x',t')\right] = -i \pi~ \text{sign}\left(t-t'-\frac{x-x'}{u}\right).\label{eq:BosonCommutation}
\end{equation}  

\begin{figure}[tb]
    \centering
    \includegraphics[width=0.9\linewidth]{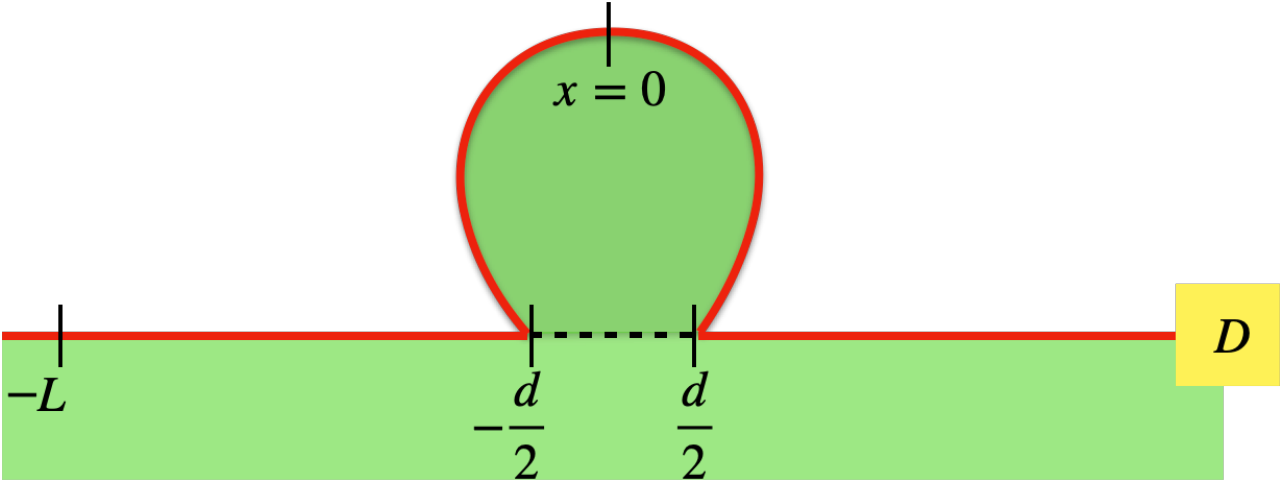}
    \caption{Schematic representation of the proposed setup: we consider a single chiral edge state (red) of a fractional quantum Hall system (green). A droplet of total perimeter $d$ is formed by applying a negative voltage to metallic gates, shaping the edge into an ``$\Omega$-junction.'' We define a curvilinear coordinate $x$ along the edge, setting $x=0$ at the center of the droplet. Tunneling processes occur within the junction, specifically between the points $x=-d/2$ and $x=d/2$. Fractional quasi-particles are injected at position $x=-L$ and propagate along the edge, while the charge current is measured at the drain, located at $x=D$ (yellow).}
    \label{fig:scheme}
\end{figure}

To investigate anyonic braiding effects in this system, we introduce a controlled deformation of the edge state, leading to the formation of a localized droplet of total perimeter $d$. This configuration, depicted in Fig.~\ref{fig:scheme}, results in the creation of an $\Omega$-shaped junction, where quasi-particles can tunnel between two distinct edge segments. We adopt a curvilinear coordinate system along the edge, with $x=0$ positioned at the center of the loop. The tunneling processes take place within the junction region, specifically between the points $x=-d/2$ and $x=d/2$. We refer to this region as the \textit{junction}, while the portion of the edge between $-d/2<x<d/2$ is designated as the \textit{loop}.  

The tunneling Hamiltonian describing quasi-particle transfer between the two sides of the junction is given by~\cite{Kane94}  
\begin{equation}
    \label{eq:HamTun}
    H_T = \Lambda \sum_{\epsilon=\pm }  e^{i\epsilon \kappa} \Psi^{\dagger}\left(\frac{\epsilon d}{2}\right)\Psi\left(-\frac{\epsilon d}{2}\right),
\end{equation}  
where $\Lambda$ represents the real part of the tunneling amplitude, while $\kappa$ accounts for a possible complex phase in the tunneling process. Here, $\epsilon = \pm$ is used as a convenient way to write the two Hermitian-conjugate contributions to the tunneling Hamiltonian.
The strength of the tunnel coupling can be modulated by applying a negative voltage to nearby metallic gates. In this work, we focus on the weak-tunneling regime, where fractional quasi-particles tunnel through the junction with a small probability. Moreover, the parameter $\kappa$ can be related to the Aharonov-Bohm phase of quasi-particle going through the loop and it can be modified by tuning the magnetic field while remaining on the plateau of the filling factor $\nu = 1/(2n+1)$.  

In Eq.~\eqref{eq:HamTun}, the quasi-particle creation and annihilation operators are expressed in terms of the bosonic field through the standard bosonization identity~\cite{vonDelft98}  
\begin{equation}
\label{eq:Anyons}
    \Psi(x) = \frac{\mathcal{F}}{\sqrt{2\pi a}} e^{-i \sqrt{\nu}\phi(x)},
\end{equation}  
where $\mathcal{F}$ is a Klein factor~\cite{Guyon2002} and $a$ is a short-distance cutoff parameter. This formalism allows us to describe the statistical properties of anyonic excitations within the fractional quantum Hall edge and investigate their braiding behavior in the $\Omega$-junction geometry.  

We consider the general case of a state with $M$ injected anyons given by
\begin{equation}
	\left| {\varphi} \right \rangle =  \frac{1}{\sqrt{\mathcal{N}}}\prod_{\mu=1}^M\Psi^{\dagger}(-L,-T-\tau_{\mu})\left|0\right\rangle
	\label{eq:AnyonState}
\end{equation}
where $\mathcal{N}$ is a normalization factor given by
\begin{equation}
\mathcal{N} = \prod_{\mu=1}^M\left\langle 0\right| \Psi(-L,-T-\tau_{\mu})\Psi^{\dagger}(-L,-T-\tau_{\mu})\left|0\right\rangle.
\label{eq:Normalization}
\end{equation}
The injection times are ordered such that
\begin{equation}
\tau_1 < \tau_{2} < \dots < \tau_{M}.
\end{equation}
These times are not necessarily equally spaced, allowing for a general sequence of injection events. We assume that the injection position is fixed at $L = uT$, where $u$ is the propagation velocity along the edge and $T$ is a reference time interval. With this choice, each quasi-particle labeled by $\mu$ arrives at the center of the loop in $x=0$ at time $\tau_{\mu}$, facilitating a clear correspondence between injection and detection times. Without loss of generality, we set the first injection time to $\tau_1 = 0$. In order to simplify the analytical treatment, we neglect the finite temporal width of the anyon pulses and model them as sharp injections. This approximation does not qualitatively affect the physical results, as the impact of finite-width pulses is known to be negligible in the case of Laughlin states, as discussed in Refs.~\cite{Ruelle23,Iyer24,Thamm24}.

The transport quantities will be computed as averages over states with multiple injected anyons like the one in Eq.~\eqref{eq:AnyonState}.

\section{Correlation functions}
\label{sec:CorrelationFunctions}
In this Section, we present the calculation for the correlation functions for the anyon operators in Eq.~\eqref{eq:Anyons} that will be used to compute the transport properties in our setup. {For} this purpose, we introduce the boson Green's function in the Keldysh formalism~\cite{Martin05} 
\begin{equation}
G^{\eta\eta'}\left(x,t;x',t'\right) \equiv \left\langle T_K\left[\phi\left(x,t \eta\right)\phi\left(x',t' \eta'\right)\right]\right\rangle-\left\langle\phi^2\left(0\right)\right\rangle.\label{eq:DefinitionBosonGreenFunction}
\end{equation}
We emphasize that, in the above equation, the average is taken over the vacuum state $\left|0\right\rangle$. Although the initial states used to compute transport quantities will involve multiple anyons, we will express the thermal averages in terms of expectation values over the vacuum. As a result, all relevant quantities can be recast in terms of the Green's functions introduced in Eq.~\eqref{eq:DefinitionBosonGreenFunction}.

The Keldysh product in Eq.~\eqref{eq:DefinitionBosonGreenFunction} can be written as
\begin{align}
& T_K[\phi\left(x, t \eta \right)\phi\left(x', t' \eta' \right)] = \nonumber \\
& \qquad \frac{1}{2}\Big[\left\{\phi\left(x,t\right),\phi\left(x',t'\right)\right\}+\sigma_{tt'}^{\eta\eta'}\left[\phi\left(x,t\right),\phi\left(x',t'\right)\right] \Big] ,
\end{align}
where $\left[\dots\right]$ is the commutator, $\left\{\dots\right\}$ is the anti-commutator and
\begin{equation}
\sigma_{tt'}^{\eta\eta'} = \text{sign}\left(t-t'\right)\frac{\eta+\eta'}{2} +  \frac{\eta'-\eta}{2},
\end{equation}
ensures the correct sign for the order along the Keldysh contour. By computing the average for a quadratic boson Hamiltonian and using the boson commutation rules defined in Eq.~\eqref{eq:BosonCommutation}, we arrive at the expression
\begin{align}
G^{\eta\eta'}\left(x,t;x',t'\right) &= \mathcal{G}\left(t-t'-\frac{x-x'}{u}\right)  \nonumber \\
& - \frac{i\pi }{2}\sigma_{tt'}^{\eta\eta'}\text{sign}\left(t-t' - \frac{x-x'}{u}\right),\label{eq:BosonCorrelator}
\end{align}
where
\begin{equation}
\mathcal{G}\left(t\right) = \ln \left|\frac{\sinh\left(i\frac{\pi}{\beta}\tau_0\right)}{\sinh\left[\frac{\pi}{\beta}\left(t+i\tau_0\right)\right]}\right|,
\end{equation}
with $\beta = k_{\mathcal{B}} \theta$ is the inverse temperature and $\tau_0 = a/u$ is a short-time cut-off. 

The correlation functions involving anyon vertex operators can be computed by means of the following generalized Wick's theorem~\cite{vonDelft98}
\begin{align}
&\left\langle \prod_j T_K\left[e^{i \alpha_j \phi\left(x_j,t_j \eta_j\right)} \right]\right\rangle = e^{-\sum_{\substack{j<l}}\alpha_j\alpha_l G^{\eta\eta'}\left(x_j,t_j;x_l,t_l\right)},\label{eq:ProductBosonCorrelator}
\end{align}
with the condition $\sum_j \alpha_j =0$.

In real samples, the edge state can couple to its electromagnetic environment, with possible dissipation effects. It can be shown that the latter effects can be taken into account by a multiplicative factor $F$ in front of the boson correlation function, such that $\mathcal{G}\left(t\right) \rightarrow F \mathcal{G}\left(t\right)$ (see, for instance, Ref.~\cite{Braggio12}). In contrast, the second contribution in Eq.~\eqref{eq:BosonCorrelator} arises due to the boson commutation relation and, therefore, it is independent of dissipation. Phenomenologically, one can recast the correlator for quasi-particle vertex operators as~\cite{Braggio12,Rosenow16}
\begin{align}
    &\left\langle T_K\left[e^{i \sqrt{\nu} \phi\left(x,t \eta\right)}e^{-i \sqrt{\nu} \phi\left(x',t' \eta'\right)} \right]\right\rangle= \nonumber \\
    & \qquad  e^{\delta \mathcal{G}\left(t-t'-\frac{x-x'}{u}\right)} e^{-\frac{i \lambda\pi }{2}\sigma_{tt'}^{\eta\eta'}\text{sign}\left(t-t' - \frac{x-x'}{u}\right)},
\end{align}
with $\delta = F\nu$ is the scaling dimension which depends on the specific sample and $\lambda$ is the statistical angle which is topologically protected and universal.

Since all Green's functions in our setup are translationally invariant in both space and time, we adopt the simplified notation
\begin{equation}
G^{-+}\left(x,t;x',t'\right) \equiv G\left(x - x', t - t'\right) \label{eq:NotationG},
\end{equation}
throughout the remainder of the paper. This notation is also justified by the fact that all transport quantities relevant to our analysis can be expressed in terms of the lesser Keldysh Green's function $G^{-+}$.

\section{Time-dependent charge current for isolated anyon pulses}
\label{sec:Current}
To illustrate the main physical mechanism at play in the proposed setup, we begin by computing the time-dependent current resulting from the injection of a state containing $M$ anyons. At this stage, the injection protocol is left unspecified in order to maintain a general and model-independent discussion.
The charge current is defined in terms of the bosonic charge density $\rho(x,t) = e\sqrt{\nu}/\left(2\pi\right)\partial_x \phi(x,t)$ as
\begin{equation}
	I(t) = e\frac{u\sqrt{\nu}}{2\pi} \left[\partial_x \phi(x,t)\right]_{x=D},
\end{equation}
where we fix the position of the detector at $x = D$. 

In the Keldysh formalism for out-of-equilibrium systems~\cite{Martin05}, we can write the average value of the current as
\begin{equation}
	\langle I(t)\rangle = \frac{eu\sqrt{\nu}}{2\pi}  \left\langle T_{K}  \left[\partial_x \phi(x,t_-)\right]_{x=D} S_K\left(-\infty,-\infty\right)\right\rangle_{{\varphi}},\label{eq:def_curr}
\end{equation}
{where we introduced the prepared state $\left| \varphi \right\rangle$ in Eq.~\eqref{eq:AnyonState}.}
We remark that the choice of the $-$ branch is arbitrary. The average over $\left| {\varphi} \right \rangle$ implies to order the operators on a $4$-branch Keldysh contour. The convention is that we set $\eta = 1$ for $\left| {\varphi} \right \rangle$, $\eta = 4$ for $\left\langle {\varphi} \right|$ and the remaining times on the "standard" part of the contour with $\eta = 2,3$~\cite{Jonckheere23}. 

To make use of the formalism developed in the previous Section—particularly the generalized Wick's theorem for computing expectation values involving products of anyon operators—we will recast the current in terms of a bosonic vertex operator:
\begin{equation}
	I(t) = \lim\limits_{\gamma \rightarrow 0} -\frac{iue\sqrt{\nu}}{2\pi \gamma} \left[\partial_x e^{i\gamma \phi(x,t)}\right]_{x=D}.
	\label{eq:GammaCurrent}
\end{equation}

\subsection{Zeroth-order calculation}
We start {with} the simplest calculation: the zero-order charge current. This will allow us to set the stage for later more involved calculations and to introduce some important definitions.
The zeroth-order current is 
\begin{equation}
	\langle I(x,t)\rangle^{(0)} = \frac{e u \sqrt{\nu}}{2\pi} \left\langle T_{K}  \left[\partial_x \phi(x,t_-)\right] \right\rangle_{{\varphi}},
\end{equation}
where the Keldysh time-evolution operator has been removed. It is convenient to express the current as
\begin{equation}
\label{eq:curr0_1}
	\langle I(x,t)\rangle^{(0)} = -\frac{i e u \sqrt{\nu}}{2\pi\gamma} \lim_{\substack{\gamma\rightarrow 0}}\partial_xK_{\gamma}^{(0)}\left(x,t\right),
\end{equation}
in terms of a product of vertex operators of boson fields
\begin{widetext}
\begin{equation}
\label{eq:corr0}
K_{\gamma}^{(0)}\left(x,t\right) = \frac{1}{\mathcal{N}}\left\langle T_{K} \left[\prod_{\mu} e^{-i \sqrt{\nu}\phi\left(-uT,-T-\tau_{\mu_4}\right)}e^{i\gamma \phi\left(x,t^-\right)}\prod_{\mu'} e^{i \sqrt{\nu}\phi\left(-uT,-T-\tau_{\mu'_1}\right)}\right]\right\rangle,
\end{equation}
\end{widetext}
{where $\mathcal{N}$ is a normalization factor introduced in Eq.~\eqref{eq:Normalization}.}
We can compute the average in Eq.~\eqref{eq:corr0} by means of the generalized Wick's theorem presented in Eq.~\eqref{eq:ProductBosonCorrelator}. One finds
\begin{align}
K_{\gamma}^{(0)}\left(x,t\right) =& \prod_{\mu}
e^{\gamma \sqrt{\nu} G \left(-uT-x,-T-\tau_{\mu}-t \right)} \nonumber \\
& \times e^{- \gamma \sqrt{\nu} G \left(uT+x,+T+\tau_{\mu}+t \right) },
\end{align}
where we used the notational prescription given in Eq.~\eqref{eq:NotationG}. Notice that the normalization factor has been canceled by an identical term appearing in the numerator. We use the property given in Eq.~\eqref{eq:UsefulRelation2} for the difference between two Green's functions with opposite time and space coordinates to write
\begin{align}
K_{\gamma}^{(0)}\left(x,t\right) = e^{i\pi \gamma \sqrt{\nu}\sum_{\mu}\text{sign}\left(t+\tau_{\mu}-\frac{x}{u}\right)},
\label{eq:corr0_computed}
\end{align}
Then, we compute the derivative with respect to $x$ and we set $\gamma = 0$, thus finding
\begin{align}
\langle I(t)\rangle^{(0)} &= -\nu e\sum_{\mu}\delta\left(\tau_{\mu}+t - \frac{D}{u}\right) .
\label{eq:curr_fin0}
\end{align}
The above expression corresponds to the current in the absence of tunneling at the junction. It describes a train of delta-function pulses, each carrying charge $-e^*$, representing the infinitely narrow anyon pulses injected at $x = L$ and propagating chirally with velocity $u$.

\subsection{First-order calculation}
The current at first order can be obtained by expanding the Keldysh time-evolution operator in Eq.~\eqref{eq:def_curr} at first order in the tunneling Hamiltonian, thus obtaining
\begin{align}
&\langle I(t)\rangle^{(1)} = \frac{e u \sqrt{\nu}}{2\pi} \frac{\Lambda}{i 2\pi a}\sum_{\substack{\eta=\pm\\\epsilon=\pm}}\eta e^{i\epsilon \kappa}\left\langle T_{K}  \left\{\left[\partial_x \phi(x,t_-)\right]_{x=D}\right.\right. \\&\times\left.\left.\int_{-\infty}^{+\infty} dt_1~ e^{i\sqrt{\nu}\phi\left(\epsilon \frac{d}{2},t_1\eta\right)}e^{-i \sqrt{\nu}\phi\left(-\epsilon\frac{d}{2},t_1\eta\right)} \right\}\right\rangle_{\Phi},
\end{align}
where $\eta$ is the index for the Keldysh summation, $\epsilon$ is an index who takes into account the Hermitian conjugate of the tunneling Hamiltonian and $\kappa$ is a phase associated with the Fermi momentum $k_F$ and the magnetic flux $\Phi$ [see Eq.~\eqref{eq:HamTun}].

Similarly to the zeroth-order case, it is useful to express the current operator in terms of a boson vertex operator using Eq.~\eqref{eq:GammaCurrent}, such that the first-order current can be recast as
\begin{align}
\label{eq:curr1_1}
		\langle I(t)\rangle^{(1)} &= -\frac{e u \sqrt{\nu}}{4\pi^2}\frac{\Lambda}{a}\sum_{\substack{\eta=\pm\\\epsilon=\pm}}\eta e^{i\epsilon \kappa} \lim_{\gamma\rightarrow 0}\frac{1}{\gamma}\left. \partial_x K_{\substack{\gamma\\ \epsilon\eta}}^{(1)}\left(x,t\right)\right|_{x=D},
\end{align}
in terms of the following correlator 
\begin{widetext}
\begin{equation}
\label{eq:corr1}
K_{\substack{\gamma\\ \epsilon\eta}}^{(1)}\left(x,t\right) = {\frac{1}{\mathcal{N}}} \Bigg\langle T_{K} \Bigg[\prod_{\mu} e^{-i \sqrt{\nu}\phi\left(-uT,-T-\tau_{\mu_4}\right)}e^{i\gamma \phi\left(x,t_{-}\right)}\int dt_1~e^{i\sqrt{\nu} \phi\left(\frac{\epsilon d}{2},t_1\eta\right)}e^{-i\sqrt{\nu} \phi\left(-\frac{\epsilon d}{2},t_1\eta\right)}\prod_{\mu'} e^{i \sqrt{\nu}\phi\left(-uT,-T-\tau_{\mu'_1}\right)}\Bigg]\Bigg\rangle.
\end{equation}
\end{widetext}
We compute the average by means of the generalized Wick's theorem~\cite{vonDelft98}
\begin{align}
&K_{\substack{\gamma\\ \epsilon\eta}}^{(1)}\left(x,t\right) = e^{\delta \mathcal{G}\left(\frac{d}{u}\right)}e^{\gamma \sqrt{\nu}\sum_{\mu}\left[G\left(-x,-t-\tau_{\mu}\right)-G\left(x,t+\tau_{\mu}\right)\right]}\nonumber\\&\times\int dt_1~ \mathcal{B}_{\epsilon}(t_1)e^{-\gamma \sqrt{\nu}\epsilon\left[G^{-\eta}\left(x,\frac{d}{2};t,t_1\right)-G^{-\eta}\left(x,-\frac{d}{2};t,t_1\right)\right]},\label{eq:corr1_computed}
\end{align}
where we used the property in Eq.~\eqref{eq:UsefulRelation4} for the Green's functions $G^{\pm\pm}$ and we also introduced the braiding term
\begin{align}
\mathcal{B}_{\epsilon}(t_1) &\equiv e^{-\lambda\epsilon\sum_{\mu}\sum_{\chi}\chi\left[G\left(\chi\frac{d}{2},-t_1-\tau_{\mu}\right)-G\left(-\chi\frac{d}{2},t_1+\tau_{\mu}\right)\right]} =\nonumber \\
&\exp\left\{i\pi\lambda\epsilon\sum_{\mu}\sum_{\chi=\pm}\chi\text{sign}\left(t_1+\tau_{\mu}-\chi\frac{d}{2u}\right)\right\},
\label{eq:BradingTerm}
\end{align} 
where the second line has been computed by exploiting the property in Eq.~\eqref{eq:UsefulRelation2}. 

The current in Eq.~\eqref{eq:curr1_1} becomes
\begin{align}
\langle I(t)\rangle^{(1)} 
=& \frac{ie^*}{2\pi}\frac{\Lambda}{a} e^{\delta \mathcal{G}\left(\frac{d}{u}\right)}\int dt_1 ~\mathcal{B}_{\epsilon}(t_1)\sum_{\substack{\eta=\pm\\\epsilon=\pm}}\eta e^{i\epsilon \kappa} \nonumber\\
&\times  \Bigg\{ \sum_{\mu}\delta\left(t+\tau_{\mu}-\frac{D}{u}\right)  \nonumber \\
& -  \frac{iu\epsilon}{2\pi}\sum_{\xi=\pm}\xi \left. \partial_{x} G^{-\eta}\left(x,\xi\frac{d}{2};t,t_1\right) \right|_{x=D} \Bigg\}.
\end{align}
We now perform the sum over $\eta$. We observe that the first term in the curly brackets is independent of $\eta$ and therefore vanishes upon performing the sum $\sum_{\eta=\pm}\eta = 0$ while the second one can be computed by using the relation in Eq.~\eqref{eq:UsefulRelation3}, such that
\begin{align}
\langle I(t)\rangle^{(1)} =ie^*\frac{\Lambda e^{\delta \mathcal{G}\left(\frac{d}{u}\right)} }{2\pi a} & \sum_{\epsilon,\xi=\pm} \epsilon\xi e^{i\epsilon \kappa} \nonumber \\
& \times \mathcal{B}_{\epsilon}\left(t-\frac{D-\xi d/2}{u}\right).
\end{align}
The sum over $\epsilon$ can also be simplified by noticing that
\begin{align}
&\sum_{\epsilon=\pm} e^{i\epsilon \kappa} \frac{\epsilon}{2i} \mathcal{B}_{\epsilon}\left(t\right) = \nonumber \\
&\qquad \sin \left\{ \kappa + \pi\lambda \sum_{\mu}\sum_{\chi=\pm}\chi~\text{sign}\left[t+\tau_{\mu}-\chi\frac{d}{2u}\right]\right\}.
\end{align}
We insert the above expression in the formula for the current, yielding
\begin{equation}
	\left\langle I(t) \right\rangle^{(1)} = I_0\sum_{\xi=\pm} \xi \sin\left\{\kappa + 2\pi \lambda \delta N\left(t-\xi \frac{d}{2u}\right)\right\},\label{eq:CurrentFinal}
\end{equation}
where
\begin{equation}
I_0 = e^*\frac{\Lambda e^{\delta\mathcal{G}\left( \frac{d}{u} \right)}}{ \pi a},
\end{equation}
is the current amplitude that contains all the non-universal parameters such as the scaling dimension and the temperature and
\begin{align}
& \delta N\left(t\right) = \nonumber \\
& \qquad \frac{1}{2} \sum_{\mu} \sum_{\chi=\pm} \chi~\text{sign}\left(t+\tau_{\mu}-\frac{D}{u}- d\frac{\chi}{2u}\right).
\label{eq:DefdeltaN}
\end{align}
is the number of anyons inside the loop at time $t$, in the out-of-equilibrium configuration. 

\subsection{Braiding processes}
Before presenting the detailed results for the time-dependent current, we elucidate the principal mechanism responsible for its emergence. To this end, we recast the charge current in terms of an average over anyons in Appendix~\ref{app:CurrentAnyons}. From Eq.~\eqref{eq:def_curr_final}, the current reads
\begin{equation}
\langle I(t) \rangle^{(1)}
= i e^* \sum_{\epsilon = \pm 1} e^{i \epsilon \kappa}
F_{\epsilon}\left(t + \epsilon \frac{\tau_d}{2} - \frac{D}{u} \right) + \text{H.c.},
\end{equation}
where we introduced for convenience the shorthand notation $\tau_d = d/u$ and the functions $F_{\pm}(t)$ are defined as
\begin{equation}
F_{\pm}(t) =
\left\langle 0 \right|
\Psi(0,0)
\Psi^{\dagger}\left(\pm \frac{d}{2}, t  \right)
\Psi\left(\mp \frac{d}{2}, t  \right)
\Psi^{\dagger}(0,0)
\left|0\right\rangle,
\end{equation}
and correspond to expectation values of four anyon operators.  These expressions are particularly useful to interpret the time-dependent current in terms of braiding processes.

To fix ideas, let us focus on $F_+(t)$. This object can be interpreted as the scalar product between two physical states:
\begin{align}
F_+(t) =&
\left[ \left\langle 0 \right| \Psi(0,0) \right] \nonumber \\
& \quad  \cdot
\left[\Psi^{\dagger}\left(\frac{d}{2}, t  \right)
\Psi\left(-\frac{d}{2}, t \right)
\Psi^{\dagger}(0,0)
\left|0\right\rangle
\right].
\label{eq:AveragePhysical}
\end{align}

\begin{figure}[tb]
\includegraphics[width=0.48\textwidth]{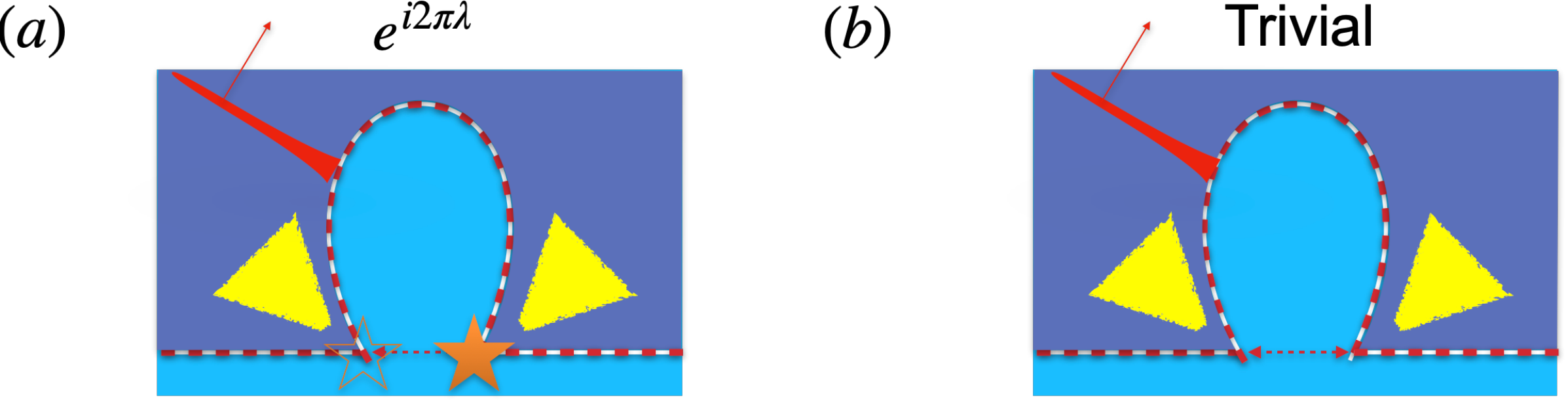}
    \caption{
Illustration of the two interfering processes responsible for the statistical phase in the time-dependent current. In both cases, an anyon (in red) impinges on the junction and enters the loop. In the panel (a), a quasi-particle--quasi-hole pair (in orange) is spontaneously created at the ends of the junction while the anyon is inside the loop, giving rise to a phase $e^{i 2\pi \lambda}$; in the other panel (b), no such pair is created.}
\label{fig:OmegaPhysical}
\end{figure}

These two states correspond to distinct physical processes differing by the spontaneous creation of a quasi-particle--quasi-hole pair at the ends of the tunneling junction. In both processes, an anyon impinges on the junction and enters the interferometer loop. In one event [the ket on the right-hand side of Eq.~\eqref{eq:AveragePhysical}] a quasi-particle-quasi-hole pair is created while the anyon is inside the loop. In the other (the bra on the left-hand side) no such pair is created. These two quantum histories, illustrated in Fig.~\ref{fig:OmegaPhysical}, can interfere. This interference is at the origin of the braiding phase observed in the current.

To make contact with the standard time-domain picture of anyonic braiding (see, for instance, Ref.~\cite{Ruelle24}), it is instructive to reinterpret the interference described above in terms of fictitious events that, while not representing real-time dynamics, better highlight the topological aspects of the process. In particular, since the current is already non-vanishing at first order in the tunneling amplitude, it can be recast as the result of the interference between two quantum amplitudes that differ by the braiding of the impinging anyon with a single quasi-particle created at the junction (specifically, at $x = -d/2$).

This interpretation is encoded in the following scalar product, which expresses the interference function $F_+(t)$:
\begin{align}
F_+(t) &=
\left[ \left\langle 0 \right| \Psi(0,0)\, \Psi^{\dagger}\left(0, t - \frac{\tau_d}{2} \right)\right] \nonumber\\
&\quad\cdot
\left[\Psi\left(0, t + \frac{\tau_d}{2} \right)
\,\Psi^{\dagger}(0,0)
\left|0\right\rangle
\right].
\label{eq:AveragePhysicalMinus}
\end{align}

\begin{figure}[tb]
    \centering
    \includegraphics[width=0.48\textwidth]{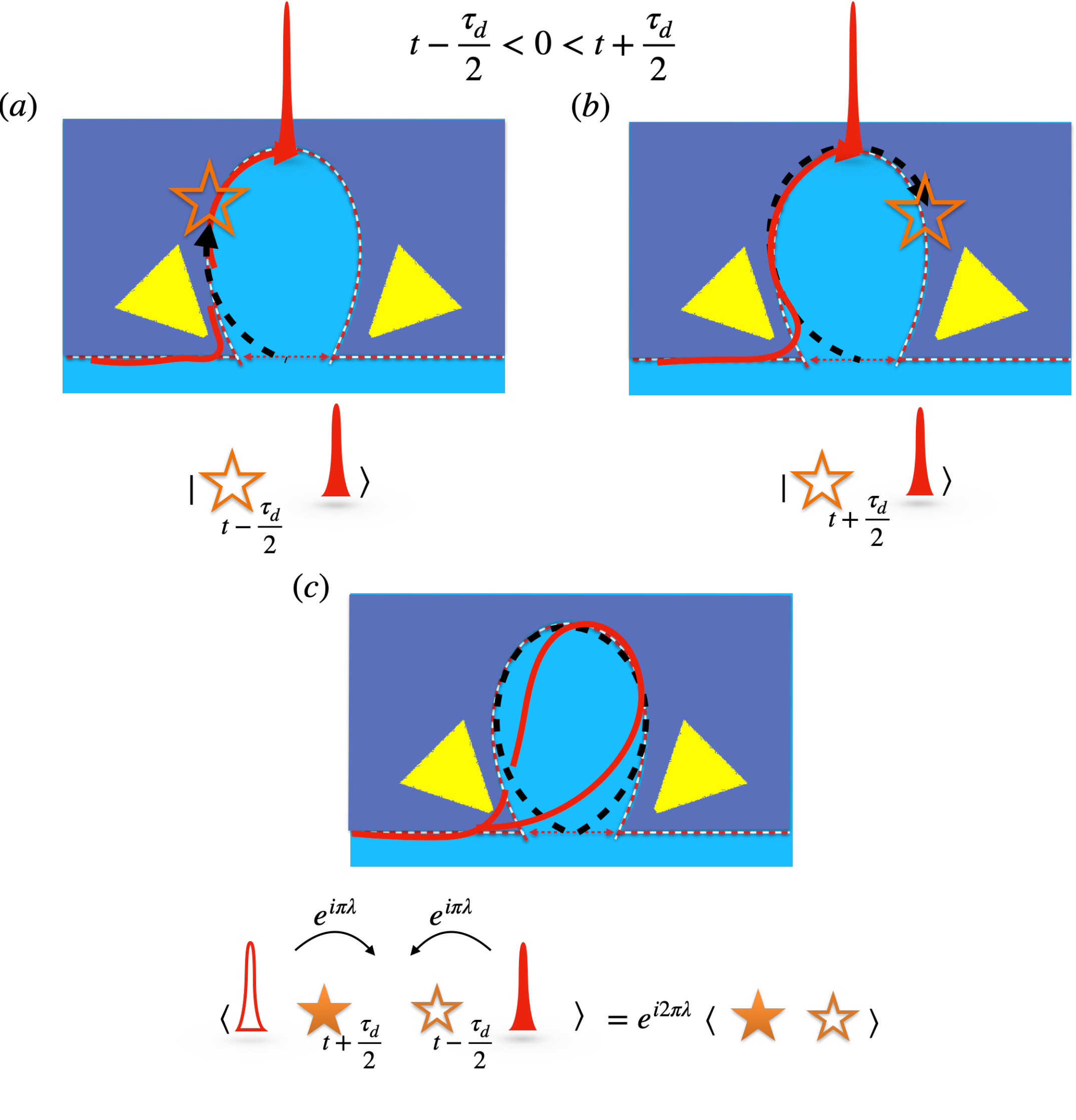}
    \caption{
    Illustration of the fundamental braiding processes responsible for the time-dependent current. The injected anyon reaches the position $x=0$ at time $t=0$. (a) The impinging anyon (red) arrives \textit{before} the quasi-hole (orange) is created at the junction, acquiring a phase \( e^{i\pi \lambda} \). (b) The quasi-hole is created \textit{before} the anyon arrives, yielding a phase \( e^{-i\pi \lambda} \). (c) The resulting current arises from the interference of these two amplitudes, leading to a net phase \( e^{i2\pi \lambda} \). Similar diagrams involving quasi-particles contribute with opposite phase, making the total current proportional to \( \sin(2\pi \lambda) \). Quasi-particles are represented by filled symbols, while quasi-holes are depicted as empty symbols with only an outline.
    }
    \label{fig:Omega1}
\end{figure}

The two corresponding processes are shown in Fig.~\ref{fig:Omega1}(a) and (b). In this picture, the situation where the impinging anyon (in red) arrives at the junction before the quasi-hole (in orange) is created acquires a statistical phase $e^{i\pi \lambda}$ due to the following exchange, occuring for $t - \tau_d/2<0$,
\begin{align}
\Psi(0,0)\, &\Psi^{\dagger}\left(0, t - \frac{\tau_d}{2} \right) =\nonumber\\&\times  e^{-i\pi \lambda \text{sign}\left(t - \frac{\tau_d}{2}\right)}\Psi^{\dagger}\left(0, t - \frac{\tau_d}{2} \right)\,\Psi(0,0).
\end{align}
Conversely, when the quasi-particle is created first, such as in the case $t + \tau_d/2>0$, the resulting phase is $e^{-i\pi \lambda}$. These configurations differ by a single exchange of positions, thus capturing the essence of time-domain braiding in our setup.

Although these events are not to be interpreted as physical time-evolution paths, they nevertheless encode the \emph{topological content} of the exchange through their mutual interference. The resulting current is determined by the scalar product between these two amplitudes [see Fig.~\ref{fig:Omega1}(c)], and carries an overall phase $e^{i2\pi \lambda}$, where the complex conjugation of $e^{-i\pi \lambda}$ accounts for its role in the ket state. Analogous processes involving a quasi-hole at $x = -d/2$ contribute with a phase $e^{-i2\pi \lambda}$. The total current is given by the difference between these two contributions, making it proportional to $\sin(2\pi \lambda)$. When $N$ anyons are present in the loop, the statistical angle generalizes to $2N\pi \lambda$.

Moreover, the spontaneous creation of quasi-particle-quasi-hole pairs can occur at the junction in two ways, meaning the process $P_1$ (where a quasi-particle is created at $x = -d/2$ and a quasi-hole at $x = +d/2$) and its reverse process $P_2$. These processes contribute to the current along the edge at two locations: $x = -d/2$ and $x = d/2$. We denote these contributions as the tunneling currents $I_T^-$ and $I_T^+$, respectively. These currents satisfy the relation $I_T^+ (t) = -I_T^-(t)$, reflecting the opposite charges of the spontaneously created anyons in both the $P_1$ and $P_2$ processes. The total current in Eq.~\eqref{eq:CurrentFinal} can be recast in terms of these currents as
\begin{equation}
\left\langle I(t) \right\rangle^{(1)} = \sum_{\xi=\pm} I_T^{\xi}\left(t-\xi \frac{\tau_d}{2}\right),
\end{equation}
with
\begin{equation}
I_T^{\xi}(t) = -I_0 \xi \sin\left\{\kappa + 2\pi \lambda \delta N\left(t\right)\right\},
\end{equation}
{where $\delta N(t)$ has been introduced in Eq.~\eqref{eq:DefdeltaN}.}

\subsection{Results for multiple anyons injection}
Here, we present the time evolution of the charge current when multiple anyons impinge on the junction. The case of a single anyon ($M=1$) has already been thoroughly discussed in Ref.~\cite{Ronetti25}. We now focus on the more complex scenario with $M>1$ anyons, choosing $M=2$ as a representative example. This minimal multi-anyon configuration already captures the key features that can be generalized to an arbitrary number of injected anyons.

\begin{figure}[tb]
\centering
    \includegraphics[width=0.48\textwidth]{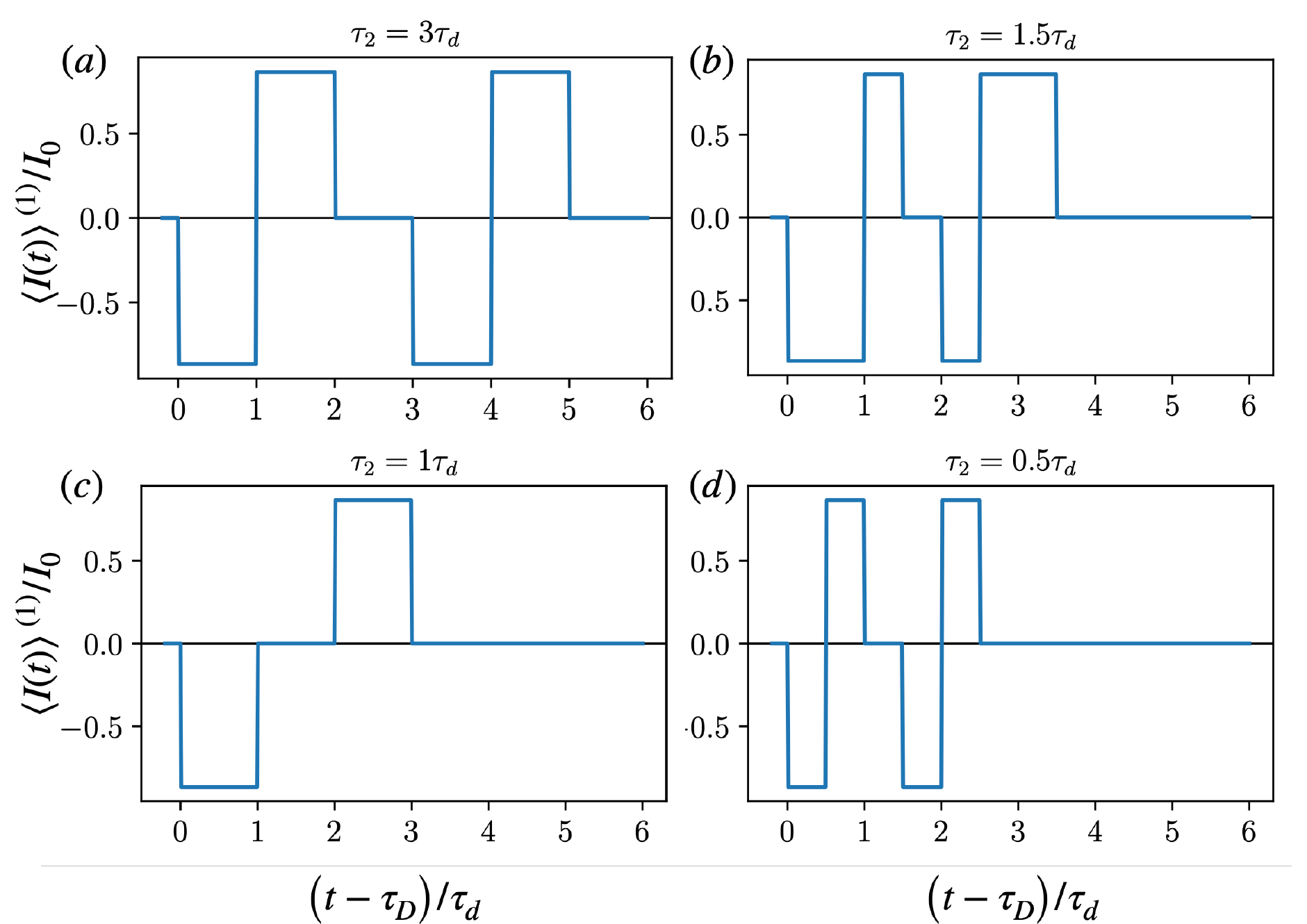}
    \caption{Time-dependent current for the case of $M = 2$ for different delays between the two anyons. Here we used the shorthand notations $\tau_d = d/u$ and $\tau_D = D/u$. The origin of time has been chosen in $t = \tau_D-\tau_d$.}
    \label{fig:TwoAnyons}
\end{figure}

Figure~\ref{fig:TwoAnyons} displays the resulting current for various separation times $\tau_2$ between the two injected anyons. In panel (a), the delay between the two injections is sufficiently large that the system exhibits two well-separated and time-shifted replicas of the single-anyon current studied in detail in Ref.~\cite{Ronetti25}. This regime corresponds to the limit of negligible overlap between the two processes.

As the separation $\tau_2$ is reduced, entering the intermediate regime $\tau_d \le \tau_2 < 2\tau_d$, shown in Fig.~\ref{fig:TwoAnyons}(b), the current signals produced by the two anyons begin to interfere. Because the contributions have opposite signs, partial cancellations occur, leading to time intervals in which the total current vanishes. This interference effect reflects the underlying anyonic statistics.

Panel (c) illustrates the special case where $\tau_2 = \tau_d$, leading to complete temporal overlap of the two steps. The resulting current profile consists of two consecutive steps of opposite sign, separated by a delay $\tau_d$. 

Finally, Fig.~\ref{fig:TwoAnyons}(d) corresponds to the regime $0 < \tau_2 < \tau_d$, where both anyons are simultaneously present in the junction at the time the quasi-particle pair is created. In this case, the interference process involves an effective braiding of the injected anyons with both components of the pair, leading to a doubling of the statistical phase from $2\pi \lambda$ to $4\pi \lambda$ over finite time intervals. For instance, when $\lambda = 1/3$, one finds $\sin\left(2\pi/3\right) = -\sin\left(4\pi/3\right)$, resulting in regions where steps of equal height alternate in sign. This behavior exemplifies the non-trivial interplay between statistics and temporal overlap in multi-anyon interference.

\section{Calculations for a dilute stream of anyons} 
\label{sec:Collider}
In the previous Section, the time-dependent charge current was computed without specifying the injection protocol for the stream of anyons. Here, we consider the emission of anyons from a source quantum point contact (QPC), drawing inspiration from pioneering works in the collider geometry~{\cite{Bartolomei20,Rosenow16,Han16,Lee22,Morel22,Mora22,Schiller23,Jonckheere23}}. A single source QPC is placed upstream of the loop junction and is operated in the weak backscattering regime. In this perturbative limit, it is well established that the tunneling excitations are anyons, whose properties are determined by the topological nature of the bulk quantum Hall state. When a constant DC bias $V_{DC}$ is applied across the source QPC, the tunneling of anyons occurs probabilistically and follows a Poissonian distribution. This configuration is schematized in Fig.~\ref{fig:SetupPoisson}. Specifically, the total number of anyons $\delta N = N-\left\langle N\right\rangle$ emitted by the source QPC during a time interval $\tau$ for $V_{DC}$ is a Poissonian random variable, whose average is given by 
\begin{equation} 
\left\langle \delta N(\tau)\right\rangle_P = \gamma \tau, 
\end{equation} 
where $-e^*\gamma$ is the average backscattering current at the QPC, which can be tuned by adjusting the value of $V_{DC}$. Above, we introduced $N$ as the number of anyons emitted by the source QPC, in the out-of-equilibrium setup, i.e when $V_{DC}\ne 0$ and $\left\langle N \right\rangle$ as the number of anyons for $V_{DC}=0$.

\subsection{Charge current}
We are now in a position to average the time-dependent charge current, computed in the previous Section, over the Poissonian distribution of the incoming anyons. Due to the chirality of the edge state, the number of anyons traversing the loop during a time interval $\tau$—as introduced in Eq.~\eqref{eq:CurrentFinal}—coincides with the number $\delta N(\tau)$ of anyons emitted at the QPC.

To perform the averaging of the time-dependent current, we need to evaluate the Poissonian average of an exponential function involving the random variable $\delta N$~\cite{Rosenow16}
\begin{equation} 
\left\langle e^{i 2\pi \lambda \left[\delta{N}\left(t_2\right)-\delta N\left(t_1\right)\right]}\right\rangle_P = e^{-\gamma \left| t_2-t_1 \right|\left(1-e^{2 i \pi \lambda ~\text{sign}\left(t_2-t_1\right)}\right)}.\label{eq:Poisson} 
\end{equation}

We now apply this result to the charge current in Eq.~\eqref{eq:CurrentFinal}, yielding 
\begin{equation} \left\langle\left\langle I(t) \right\rangle^{(1)}\right\rangle_P = I_0 \left[ \frac{e^{i\kappa}}{2i} e^{-\gamma \tau_d\left(1-e^{-i2\pi\lambda}\right)} + \text{H.c.}\right]\sum_{\xi}\xi = 0, \label{eq:CurrentCollider2} \end{equation}
due to the sum over $\xi$.

Since this contribution averages to zero, it does not provide sufficient information to extract the statistical angle. Therefore, we must turn to a different transport observable to achieve this goal.

\subsection{Finite-frequency current cross-correlations}

\begin{figure}[tb]
\centering
    \includegraphics[width=0.48\textwidth]{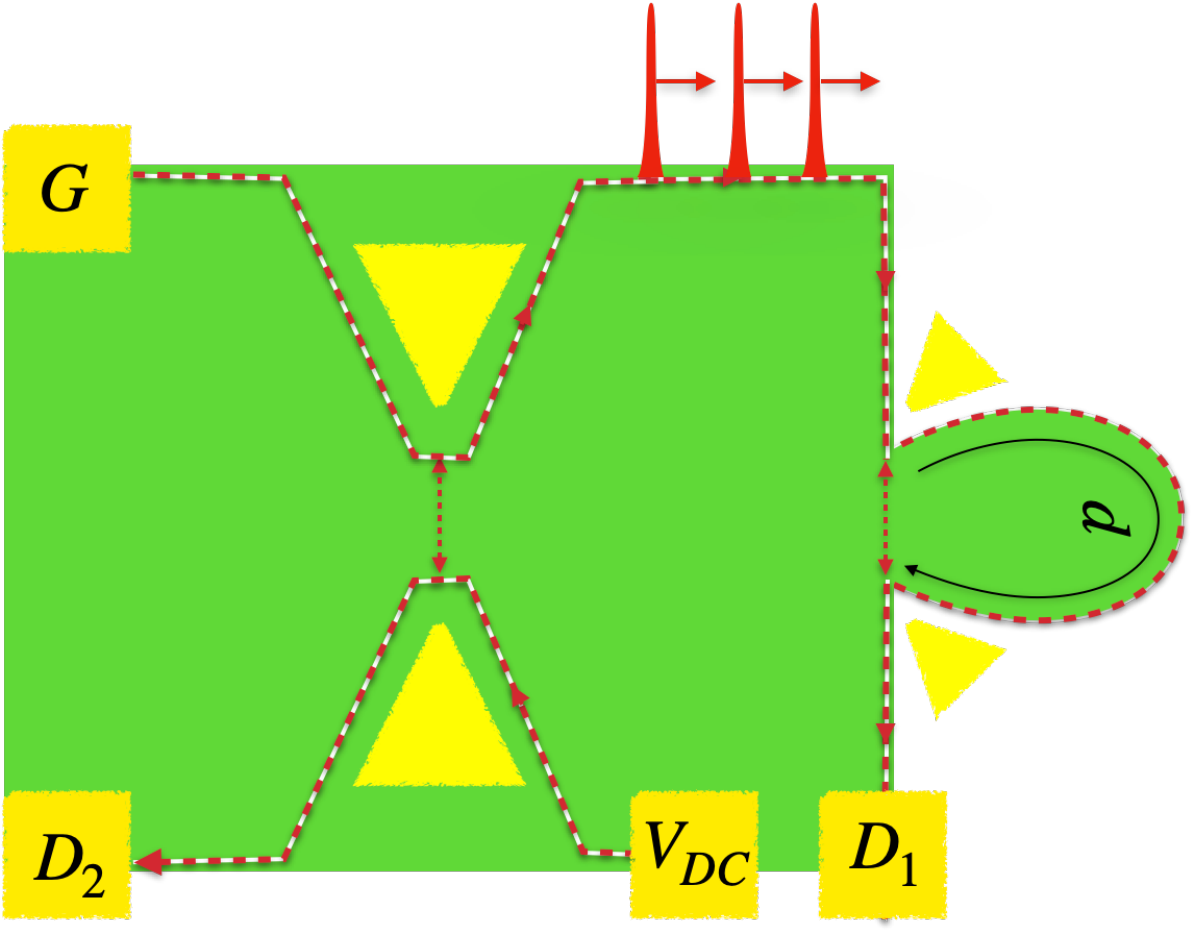}
    \caption{Schematic representation of the setup used to compute the finite-frequency cross-correlation noise between reservoirs \( D_1 \) and \( D_2 \). The system consists of two chiral edges, described by bosonic fields \( \phi_R \) and \( \phi_L \), which are coupled via a source quantum point contact (QPC) operating in the weak backscattering regime. This QPC emits a dilute stream of anyons whose number follows a Poissonian distribution. The reservoir \( D_1 \) is located downstream of the \( \Omega \)-shaped tunneling junction, and the current flowing into it encodes information about the braiding processes occurring at the junction.
    }
    \label{fig:SetupPoisson}
\end{figure}

We compute the finite frequency cross-correlation noise between reservoirs $D_1$ and $D_2$ {(see Fig.~\ref{fig:SetupPoisson} for a sketch of the setup)}. In order to take into account the effect of the source QPC, we do not average over states containing anyons but we employ the non-equilibrium bosonization formalism as in Ref.~\cite{Rosenow16}. In order to perform this calculation, we have to take into account the presence of both edges, i.e. we will have to deal with two boson fields $\phi_R$ and $\phi_L$. Moreover, all the averages will include simultaneously the thermal average and a Poissonian average over the number of anyons emitted by the source QPC. In this approach, the boson fields at positions downstream with respect to the QPC and in the absence of the $\Omega$ junction, are given by
\begin{align}
\phi_R(x,t) =& \phi_R^{(0)}(x,t) + 2\pi \frac{\lambda}{\sqrt{\nu}} \delta N\left(t-\frac{x}{u}\right),\\
\phi_L(x,t) =& \phi_L^{(0)}(x,t) + 2\pi \frac{\lambda}{\sqrt{\nu}} \delta N\left(t+\frac{x}{u}\right) \nonumber \\
& - e\sqrt{\nu} V_{DC} \left(t+\frac{x}{u}\right),\label{eq:BosonNonEqL}
\end{align}
where $V_{DC}$ is the constant DC source voltage applied across the QPC. We will neglect the last contribution in Eq.~\eqref{eq:BosonNonEqL} associated with $V_{DC}$, as it does not affect the noise calculation. The main role of this DC bias is to control the parameter $\gamma$, which is related to the current emitted from the source QPC, {with $\gamma = I_\text{source}/e^* = \mathcal{T}_S G_0 V_{DC}/e^*$, $\mathcal{T}_S$ being the transmission of the source QPC and $G_0$ the conductance along the edge}. Moreover, the reason why we omit the presence of the $\Omega$ junction is that we will introduce it perturbatively later at lowest order in the tunneling amplitude. Here, $\phi_{R/L}^{(0)}$ are the equilibrium boson fields, in the absence of both source QPC and $\Omega$ junction.

The finite-frequency cross-correlation noise is defined as
\begin{widetext}
\begin{align}
\mathcal{S}\left(\omega\right) = \int d\left(t-t'\right)e^{i \omega \left(t-t'\right)}\left\{\left\langle T_K\left[\delta I_1(t^-)\delta I_2(t'{}^+)e^{-i \int_K d\tau~H_T(\tau)}\right]\right\rangle\right\}\label{eq:defNoise}
\end{align}
\end{widetext}
where we introduced the current operator entering the two detectors $D_1$ and $D_2$
\begin{align}
I_1(t) &= \frac{e u \sqrt{\nu}}{2\pi}\partial_x\phi_R(x,t)_{x=D_1},\label{eq:CurrentSM1}\\
I_2(t) &= \frac{e u \sqrt{\nu}}{2\pi}\partial_x\phi_L(x,t)_{x=-D_2},\label{eq:CurrentSM2}
\end{align}
and their corresponding fluctuations ($j=1,2$)
\begin{equation}
\delta I_j(t) = I_j(t) - \left\langle T_K\left[ I_j(t^-) e^{-i \int_K d\tau~H_T(\tau)}\right]\right\rangle \, .
\end{equation}
We recall that the tunneling Hamiltonian $H_T$ depends only on the right-moving boson field. The average in Eq.~\eqref{eq:defNoise} is at the same time a thermal average (for the boson fields $\phi_{R/L}^{(0)}$) and a Poissonian average (for the anyon number $N$). These two averages are independent.

The current operators can be divided into two contributions as ($j=1,2$)
\begin{equation}
I_j(t) = I^{(0)}_j(t) + I^{(V)}_j(t),
\end{equation}
where
\begin{equation}
I^{(0)}_{1/2}(t) = \frac{e u \sqrt{\nu}}{2\pi}\left[\partial_x\phi^{(0)}_{R/L}(x,t)\right]_{x=D_{1/2}},
\label{eq:BareCurrentCollider}
\end{equation}
are the bare current operators in the absence of the DC bias at the QPC and 
\begin{equation}
I^{(V)}_{1/2}(t) = \mp e \lambda \partial_t\delta N\left(t\mp \frac{D_{1,2}}{u}\right),
\end{equation}
are the excess contributions in the presence of the DC bias $V_{DC}$ applied to the source QPC.

At lowest order in tunneling, the finite-frequency noise becomes
\begin{align}
\mathcal{S}^{(1)}\left(\omega\right) &= -i\int d\left(t-t'\right)e^{i \omega \left(t-t'\right)} \nonumber \\
&\times\left\langle T_K\left[I_1(t^-)I_2(t'{}^+)\sum_{\eta=\pm}\eta\int dt_1~H_T(t_1^{\eta})\right]\right\rangle \label{eq:defNoise1}
\end{align}
since at lowest order, the average of the current $I_1$ is zero for the collider geometry, see Eq.~\eqref{eq:CurrentCollider2}. 

The full average in Eq.~\eqref{eq:defNoise1} can be computed by breaking it down into the independent thermal averages on the right-moving and left-moving bosons and the independent Poissonian average on the number of injected anyons. In particular for the thermal averages, one can see that each average over bare currents alone gives
\begin{equation}
\left\langle I^{(0)}_{1/2}(t)\right\rangle = \frac{e u \sqrt{\nu}}{2\pi}\left[\partial_x\left\langle\phi^{(0)}_{R/L}(x,t)\right\rangle\right]_{x=D_{1/2}} = 0,
\end{equation}
and that the average over the tunneling Hamiltonian is
\begin{equation}
\sum_{\eta=\pm}\eta\left\langle T_K\left[H_T(t_1^{\eta})\right]\right\rangle = \sum_{\eta=\pm}\eta\left\langle T_K\left[H_T(0)\right]\right\rangle = 0,
\end{equation}
since the expression in the integral over $t_1$ becomes independent of $\eta$ after the average. Moreover, we recall that the tunneling Hamiltonian contains only right-moving bosons and, therefore, each thermal average including only the tunneling Hamiltonian and left-moving bosons vanishes. After the above relations are taken into account, the full average breaks down to 
\begin{align}
\mathcal{S}^{(1)}\left(\omega\right) &= -i\int d\left(t-t'\right)e^{i \omega \left(t-t'\right)}\int dt_1~\nonumber\\
&\times\left\langle  T_K\left[I^{(0)}_1(t^-) \sum_{\eta=\pm}\eta~H_T(t_1^{\eta})\right] I_2^{(V)}(t') \right\rangle, \label{eq:defNoise2}
\end{align}
where the average is still on both the thermal distribution of right-moving bosons and the Poissonian distribution of anyons emitted at the source QPC. The operator $I_2^{(V)}(t')$ has been taken out of the Keldysh ordering since it does not contain boson operators and it does not need to be ordered over the Keldysh contour. 

The tunneling Hamiltonian $H_T$ contains both boson operators and anyon number operators. In order to properly compute the thermal and Poissonian averages, we can recast the tunneling Hamiltonian as follows
\begin{equation}
H_T(t) = \sum_{\epsilon=\pm }e^{i\epsilon \kappa} \mathcal{B}_{\epsilon}(t)H^{(0)}_{T,\epsilon}(t),
\end{equation}
where we defined the braiding factor
\begin{equation}
\mathcal{B}_{\epsilon}(t) = \exp\left\{i 2\pi \lambda \epsilon
{\sum_{\chi=\pm}\chi ~ \delta N\left(t - \chi \frac{d}{2u} \right)}
\right\},
\label{eq:DefinitionBraidingFactor}
\end{equation}
and the bare tunneling Hamiltonian
\begin{equation}
H^{(0)}_{T,\epsilon}(t) = \frac{\Lambda}{2\pi a}e^{i\sqrt{\nu}{\phi_R}^{(0)}\left(\epsilon \frac{d}{2},t\right)}e^{-i \sqrt{\nu}{\phi_R}^{(0)}\left(-\epsilon\frac{d}{2},t\right)}.
\label{eq:BareTunnelingCollider}
\end{equation}

\begin{figure}[tb]
    \centering
    \includegraphics[width=0.95\linewidth]{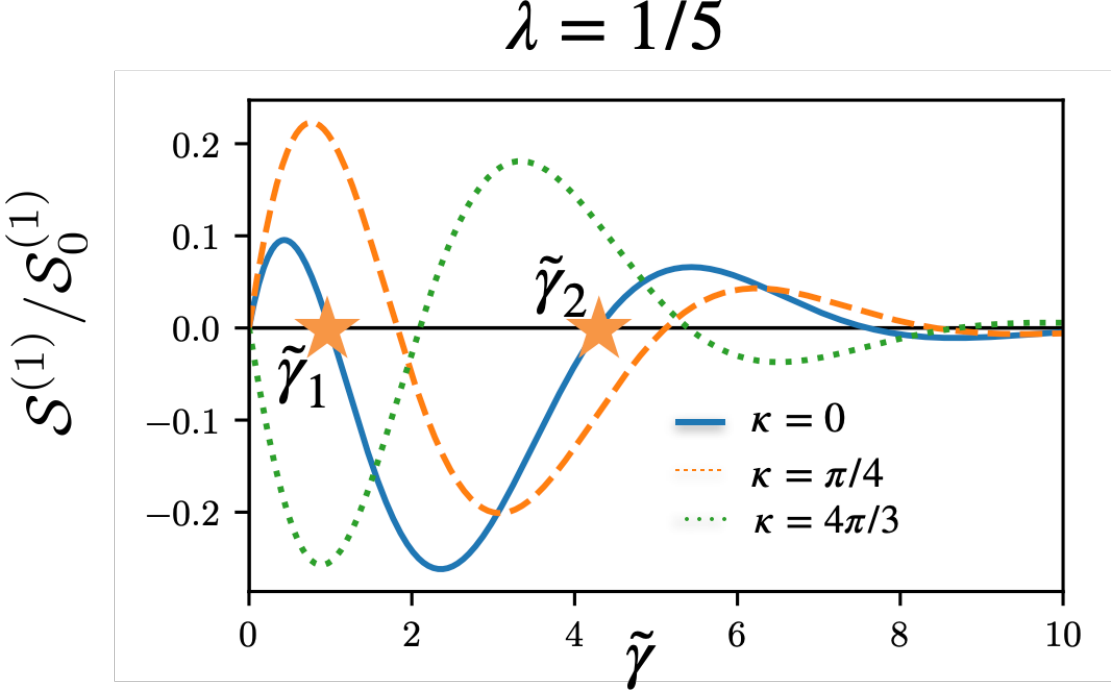}
    \caption{The finite-frequency noise $\mathcal{S}^{(1)}/\mathcal{S}^{(1)}_0$ as a function of $\tilde{\gamma}$ for $\lambda = 1/5 $ and various values of $\kappa$. From the position of the zeros $\tilde{\gamma}_1$ and $\tilde{\gamma}_2$, one can obtain the statistical angle $\lambda$. }
    \label{Fig:Noise1}
\end{figure}

The thermal average included in the full average of Eq.~\eqref{eq:defNoise2} gives
\begin{align}
&\left\langle  T_K\left[I^{(0)}_1(t^-) \sum_{\eta=\pm}\eta~H^{(0)}_{T,\epsilon}(t_1^{\eta})\right] \right\rangle = \nonumber \\
& \qquad  -\frac{e^*\Lambda e^{\delta  \mathcal{G}\left(\frac{d}{u}\right)}}{2\pi a} \epsilon \sum_{\xi=\pm } \xi \delta\left(t-t_1 -\frac{D_1-\xi d/2}{u}\right) .\label{eq:ThermalAverageCollider}
\end{align}
By plugging the above result into Eq.~\eqref{eq:defNoise2}, the expression for the finite-frequency cross-correlation noise at lowest order becomes
\begin{equation}
\mathcal{S}^{(1)}\left(\omega\right) =\frac{e^2\nu\lambda \Lambda}{\pi a} e^{\delta \mathcal{G}\left(\frac{d}{u}\right)} e^{i\omega \left(\frac{D_1-D_2}{u} \right)  }\sin \left( \omega \frac{\tau_d}{2} \right)\mathcal{B}\left(\omega\right),
\label{eq:noiseSMfin1}
\end{equation}
where we defined
\begin{equation}
\mathcal{B}\left(\omega\right) = \int d\left(t-t'\right)e^{i\omega\left(t-t'\right)}\sum_{\epsilon}\epsilon e^{i\epsilon \kappa}\left\langle \mathcal{B}_{\epsilon}(t) {\partial_{t'} \delta N \left(t'\right)} \right\rangle.\label{eq:DefinitionBetaOmega}
\end{equation}
The details of the calculation of the above quantity are given in the Appendix~\ref{app:DetailsNoise}. By inserting it {into} the expression for the finite-frequency correlation noise and taking only the real part, one is left with
\begin{align}
\text{Re} \left[\mathcal{S}^{(1)}\left(\omega\right) \right]  \equiv \mathcal{S}_0^{(1)}\left(\omega\right)\mathcal{C}_{\lambda}\left(\gamma\tau_d\right),
\label{eq:noiseSMfin2real}
\end{align}
where the amplitude for the noise is
\begin{align}
    \mathcal{S}_0^{(1)}\left(\omega\right) =& 4 \frac{e^2\nu^2 \Lambda}{\pi a} e^{\delta \mathcal{G}\left(d/u\right)} \sin \left(\omega \frac{D_1-D_2}{u} \right)  \frac{ \sin^2 \left( \omega \frac{\tau_d}{2} \right)}{ \omega \frac{\tau_d}{2}},
\end{align}
and the function which includes all the dependence on the injected current $\gamma$ is
\begin{align}
    \mathcal{C}_\lambda (\tilde{\gamma}) &= \sin (\pi \lambda)\tilde{\gamma} e^{-\tilde{\gamma} \left[1-\cos\left({2\pi \lambda}\right)\right]}\nonumber\\& \times\cos \left[\tilde{\gamma} \sin\left(2\pi\lambda\right)+\pi\lambda - \kappa\left(\Phi\right)\right].\label{eq:CrossCorrelations}
\end{align}
The noise is presented in Fig.~\ref{Fig:Noise1} for the case $\nu=1/5$ (for other fractional fillings, see Ref.~\cite{Ronetti25}). {This plot presents the finite-frequency noise for different values of $\kappa$, rescaled with respect to $\mathcal{S}_0^{(1)}$. Therefore, it corresponds to~\eqref{eq:CrossCorrelations}, which still retains a dependence on the non-universal parameter $\kappa_{\lambda}$. Nevertheless, the spacing between two successive zeros of the sine function, denoted by $\tilde{\gamma}_1$ and $\tilde{\gamma}_2$, remains unaffected by the precise value of $\tilde{\kappa}_{\lambda}$. This universal property allows for the extraction of the statistical angle $\lambda$ based solely on the measurement of such spacing. As an illustrative case, Fig.~\ref{Fig:Noise1} displays the noise profile for $\kappa = 0$, from which $\lambda$ can be determined using the simple inversion formula:
\begin{equation}
\lambda = \frac{1}{2\pi}\arcsin\left(\frac{\pi}{\tilde{\gamma}_2-\tilde{\gamma}_1}\right).
\label{eq:Lambda}
\end{equation}
This approach enables a direct determination of $\lambda$ from experimental data, bypassing the need for any a priori knowledge of non-universal system-specific parameters, such as the scaling dimension. In practical terms, one can measure the noise as a function of the dimensionless variable $\gamma$, which depends only on controllable experimental parameters such as the applied bias and the QPC transmission. The timescale $\tau_d$, entering the relation $\tilde{\gamma} = \gamma \tau_d$, can then be jointly inferred with $\lambda$ by combining Eq.~\eqref{eq:Lambda} and the observed exponential envelope of the noise signal, which decays as $\tilde{\gamma}(1 - \cos(2\pi \lambda))$.
}

\begin{figure}[tb]
    \centering
    \includegraphics[width=0.95\linewidth]{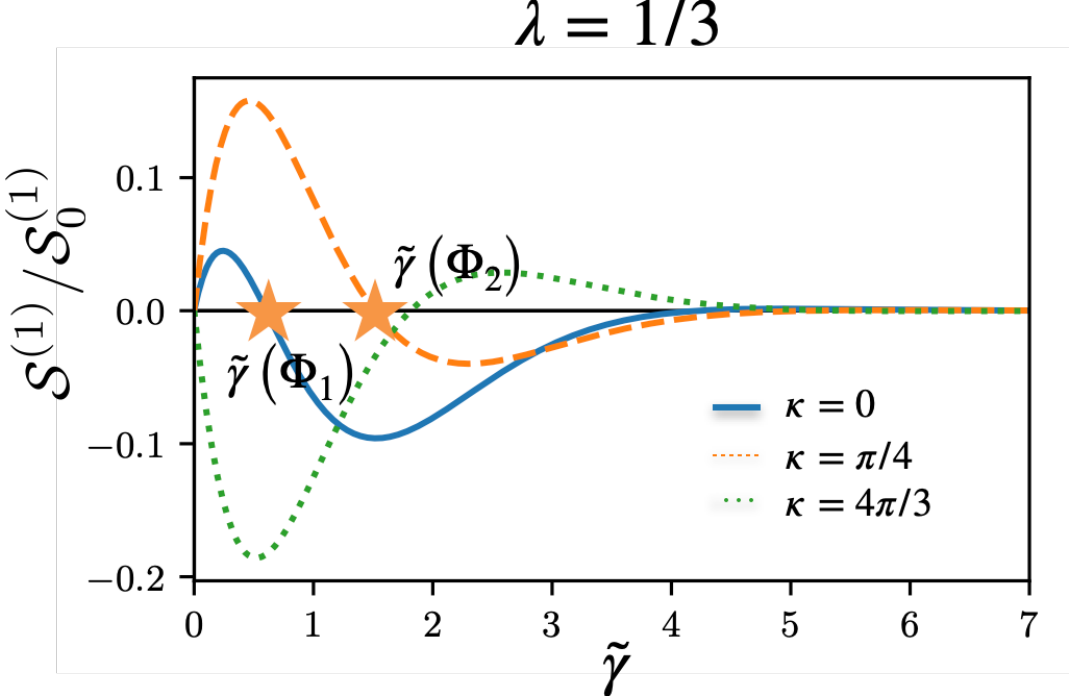}
    \caption{The finite-frequency noise $\mathcal{S}^{(1)}/\mathcal{S}^{(1)}_0$ as a function of $\tilde{\gamma}$ for $\lambda = 1/3 $ and various values of $\kappa$. From the position of the zeros $\tilde{\gamma}\left(\Phi_1\right)$ and $\tilde{\gamma}\left(\Phi_2\right)$, one can obtain the statistical angle $\lambda$ (see Eq.~\eqref{eq:Lambda2}). }
    \label{Fig:Noise2}
\end{figure}

Furthermore, we note that the value of $\kappa$ can be tuned by varying the magnetic field while remaining within the same fractional quantum Hall plateau. This allows us to select a value of $\kappa$ where the slope at the second zero is maximized. Alternatively, one can consider the difference between two zeros $\tilde{\gamma} \left(\Phi_{1/2}\right)$ at different magnetic fluxes $\Phi_1$ and $\Phi_2$ for the case of $\nu=1/3$, leading to different values of $\kappa$, as presented in Fig.~\ref{Fig:Noise2}, such that
\begin{equation}
    \tilde{\gamma}(\Phi_{1/2}) \sin(2\pi \lambda) - \tilde{\kappa}_{\lambda}(\Phi_{1/2}) = 0,
\end{equation}
where $\tilde{\kappa}_{\lambda}(\Phi) = \tilde{\kappa}_{\lambda} + 2\pi \Phi/\phi_0$, with $\Phi$ being the magnetic flux associated with $B$ and $\Phi_0 = h/e$. From this relation, $\lambda$ can be extracted as
\begin{equation}
    \lambda = \frac{1}{2\pi}\arcsin\left[\frac{2\pi}{\Phi_0}\left(\frac{\Phi_1 - \Phi_2}{\tilde{\gamma}(\Phi_1) - \tilde{\gamma}(\Phi_2)}\right)\right].\label{eq:Lambda2}
\end{equation}
Since the difference $\Phi_1 - \Phi_2$ is known, this approach provides a method for extracting the statistical angle $\lambda$.
\section{Conclusion}
\label{sec:Conclusions}
In this work, we have proposed and analyzed a novel setup aimed at extracting the statistical angle $\pi \lambda$ of topological anyons emitted by a quantum point contact, without requiring prior knowledge of the scaling dimension $\delta$. Our design is based on a single edge of a fractional quantum Hall liquid at filling factor $\nu = 1/(2n+1)$, incorporating an $\Omega$-shaped junction realized by geometrically defining a localized droplet within the FQHE fluid. This configuration draws inspiration from mesoscopic capacitor experiments in the integer quantum Hall regime.

We have shown that the time-dependent charge current generated in response to isolated anyon pulses is dominantly governed by exchange statistics. In particular, we have identified the braiding mechanisms that lead to a measurable signal, and demonstrated how the temporal features of the current encode the statistical angle, while non-universal parameters such as temperature or $\delta$ only renormalize the amplitude.

Finally, we have extended our analysis to a more experimentally accessible regime resembling an anyon collider. In this geometry, a biased QPC emits a dilute anyon stream, and we compute the finite-frequency current cross-correlations at the outputs of the $\Omega$-junction. We find that these correlations provide a robust and independent method to extract $\pi \lambda$, without requiring a separate measurement of $\delta$. The scaling dimension can instead be determined by locating two zeros of the noise as a function of the bias voltage or the magnetic flux threading the loop.

Our results demonstrate that the proposed protocol offers a promising route toward a direct detection of anyonic statistics in FQHE systems, and may pave the way for future implementations of topologically protected operations based on controlled anyon manipulation.

\acknowledgments{M. H. would like to thank Victor Bastidas for fruitful discussions at the very first stage of this work. This work was carried out in the framework of the project ``ANY-HALL" (ANR Grant No.
ANR-21-CE30-0064-03). It received support from the French government under the France 2030 investment plan, as part of the Initiative d’Excellence d’Aix-Marseille Université A*MIDEX. We acknowledge support from the institutes IPhU (AMX-19-IET008) and AMUtech (AMX-19-IET-01X). This work has benefited from State aid managed by the Agence Nationale de la Recherche under the France 2030 programme, reference ``ANR-22-PETQ-0012".
This French-Japanese collaboration is supported by the CNRS International Research Project ``Excitations in Correlated Electron Systems driven in the GigaHertz range" (ESEC).}

\appendix
\section{Properties of the Green's functions}
\label{app:PropertiesGF}
In this Appendix, we present some useful relations that are employed to simplify the correlation functions in the calculation of transport quantities. First of all, we notice that the boson Green's function in Eq.~\eqref{eq:BosonCorrelator} can be recast as
\begin{align}
G^{\eta\eta'}(x,t;x',t') &= \mathcal{G}\left(t-t'-\frac{x-x'}{u}\right) \nonumber \\
& - \frac{i\pi}{2}\left[\eta'\Theta(t-t')-\eta\Theta(-t+t')\right] \nonumber \\
& \times \text{sign}\left(t-t'- \frac{x-x'}{u}\right).\label{eq:GF}
\end{align}
From the above equation, we deduce some useful relations. First of all, we notice that for the case $\eta_1=\eta_2\equiv \eta$, we obtain at equal time $t=t'$
\begin{equation}
G^{\eta\eta}\left(x,t;x',t\right) = \mathcal{G}\left(\frac{x'-x}{u}\right),\label{eq:UsefulRelation1}
\end{equation}
since $\text{sign}(0)=0$. In Eq.~\eqref{eq:corr1_computed}, one can use the relation in Eq.~\eqref{eq:UsefulRelation1} such that
\begin{equation}
G^{\eta\eta}(\epsilon d,0) = \mathcal{G}\left(\epsilon\frac{d}{u}\right),\label{eq:UsefulRelation1b}
\end{equation}
since the Keldysh Green's function for equal index and at same time become independent of $\eta$. Moreover, one has that 
\begin{equation}
\mathcal{G}\left(t\right) = \mathcal{G}\left(-t\right),
\end{equation}
such that, we can drop the $\epsilon$ inside the argument and write
\begin{equation}
G^{\eta\eta}(\epsilon d,0) = \mathcal{G}\left(\frac{d}{u}\right).\label{eq:UsefulRelation4}
\end{equation}

Moreover, for the case $\eta_1=-\eta_2\equiv \eta$, one has
\begin{equation}
G^{\eta-\eta}(x,t;x',t') = G^{-\eta\eta}(-x,-t;-x',-t'),
\end{equation}
from which we deduce for the case $\eta=-$
\begin{align}
&G^{-+}(x,t;x',t') - G^{-+}(-x,-t;-x',-t') \nonumber \\
& \qquad = -i\pi \text{sign}\left(t-t' - \frac{x-x'}{u}\right).
\label{eq:UsefulRelation2}
\end{align}
Finally, we compute explicitly the derivative of the boson Green's function
\begin{align}
\partial_xG^{\eta_1\eta_2}(x,t) &= \partial_x\mathcal{G}\left(t- \frac{x}{u}\right) \nonumber \\
&+\frac{i\pi}{u}\left[\eta_2\Theta(x)-\eta_1\Theta(-x  )\right]\delta\left(t-\frac{x}{u}\right).
\label{eq:UsefulRelation3}
\end{align}

\section{Current in the integer quantum Hall regime: scattering matrix approach}
In this Appendix, we provide the detailed derivation of the average current at first order in the tunneling amplitude $s$ at integer filling factor $\nu=1$. We employ the scattering matrix formalism and the expression of the electronic state in energy representation to obtain an analytical expression for the current~\cite{Lesovik11}. This calculation confirms that, at filling factor $\nu=1$, the temperature-dependent contribution factorizes from the part of the current governed by the external drive, and that the current vanishes in the case of a delta-like pulse, in agreement with the absence of anyonic braiding.

\begin{figure}[tb]
\centering
\includegraphics[width=0.4\textwidth]{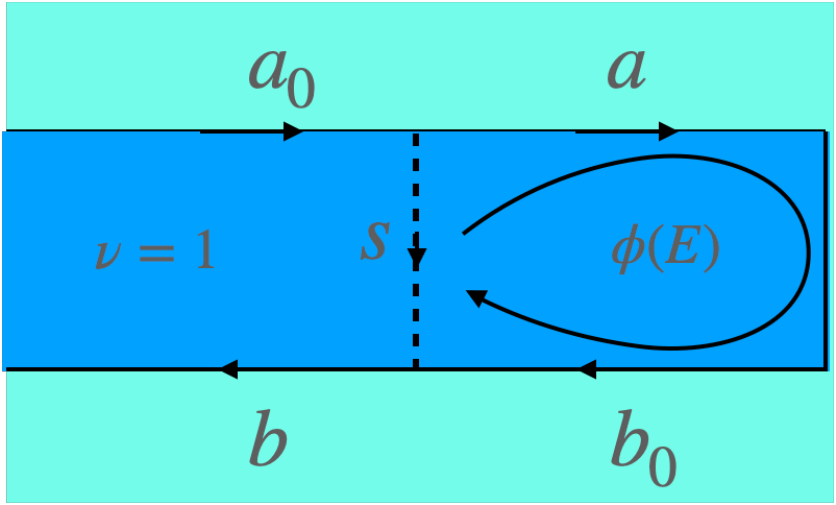}
\caption{Scattering description of the $\Omega$-junction at $\nu = 1$. The tunneling at the junction is characterized by the amplitude $s$ within the scattering matrix formalism. Incoming modes $a_0$ and $b_0$ are scattered into outgoing modes $a$ and $b$.}
\label{fig:Scattering}
\end{figure}

We compute the current coming out of the $\Omega-$junction for an edge state of the quantum Hall effect at filling factor $\nu = 1$. The scattering problem at the junction can be described in terms of four operators $a_0(E),b_0(E),a(E),b(E)$, as described in Fig.~\ref{fig:Scattering}, with the boundary condition
\begin{equation}
b_0(E) = e^{i \phi(E)} a(E),\label{eq:Boundary}
\end{equation}
where $\phi(E) = Ed/u + \phi_0$ is the phase accumulated by one electron going around the junction. These four modes are related as follows by a scattering matrix
\begin{equation}
\left(\begin{matrix}
a(E) \\b(E)
\end{matrix}\right) = \left(\begin{matrix}
r & s'\\ s & r'
\end{matrix}\right)\left(\begin{matrix}
a_0(E) \\b_0(E)
\end{matrix}\right),
\end{equation}
with the additional condition $rr' - ss' = 1$. We can solve the above system of equations, together with Eq.~\eqref{eq:Boundary}, for $b(E)$, thus finding
\begin{equation}
b(E) = S(E) a_0(E),\label{eq:ScatteringMatrix}
\end{equation}
with
\begin{equation}
S(E) = \frac{s+rr'e^{i \phi(E)}}{1-s' e^{i\phi(E)}} .
\end{equation}
We parametrize the scattering matrix such that $s = -s'= \sqrt{T}$, where $T$ is the transmission probability for an electron to pass through the junction without going around the path of length $d$. One finds
\begin{equation}
b(E) = \frac{s+\left(1-s^2\right)e^{i \phi(E)}}{1+s e^{i\phi(E)}} a_0(E).
\end{equation}
In order to take into account the presence of electrons injected on the edge states we adopt a prescription which is equivalent to the formalism employed in the main text. Instead of averaging over a state including one electron, we apply an isolated pulse $V(t) = V_0 \delta (t)$, such that 
\begin{equation}
a_0(t) \rightarrow a_0(t) e^{-i e\int_{-\infty}^t d\tau V\left(\tau\right)}.
\end{equation}
Then, in energy space one has
\begin{equation}
a_0(E) = \int dt e^{-i E t}a_0(t) = \int dE' P\left(E'\right) \tilde{a}_0\left(E-E'\right),
\end{equation}
where the operators $\tilde{a}_0\left(E\right)$ satisfy
\begin{equation}
\left\langle \tilde{a}^{\dagger}_0\left(E\right)\tilde{a}_0\left(E'\right)\right\rangle = \delta\left(E-E'\right)f_{\mu}(E),
\end{equation}
with $f_{\mu}(E)$ the Fermi function with chemical potential $\mu$ and
\begin{equation}
P(E) = \int dt e^{-i E t} e^{-i e\int_{-\infty}^t d\tau V\left(\tau\right)}.
\label{eq:defP}
\end{equation}
We can also write the average values of operator $a_0\left(E\right)$ as
\begin{align}
\left\langle a^{\dagger}_0\left(E_1\right)a_0\left(E_2\right)\right\rangle =& \int dE' P^*\left(E'\right) P\left(E'+E_2-E_1\right) \nonumber \\
&\times f_{\mu}\left(E_1-E'\right).\label{eq:Averagea0}
\end{align}

The current operator is defined as
\begin{equation}
\left\langle I(t)\right\rangle = -\frac{e}{2\pi} \left[\left\langle b^{\dagger}(t)b(t)\right\rangle - \left\langle a_0^{\dagger}(t)a_0(t)\right\rangle\right].
\end{equation}
Transforming to the energy representation, we obtain
\begin{align}
\left\langle I(t)\right\rangle =& -\frac{e}{2\pi} \int dE\, dE'\, e^{-i\left(E-E'\right)t}
\nonumber \\
&\times\left[\left\langle b^{\dagger}(E)b(E')\right\rangle - \left\langle a_0^{\dagger}(E)a_0(E')\right\rangle\right].
\end{align}
Using the scattering relation in Eq.~\eqref{eq:ScatteringMatrix}, the expression becomes
\begin{align}
\left\langle I(t)\right\rangle = & -\frac{e}{2\pi} \int dE\, dE'\, e^{-i\left(E-E'\right)t}
\left[S^*(E)S(E') - 1 \right] \nonumber \\
&\times \left\langle a_0^{\dagger}(E)a_0(E')\right\rangle.
\end{align}
Substituting Eq.~\eqref{eq:Averagea0}, we find
\begin{align}
\left\langle I(t)\right\rangle 
&= -\frac{e}{2\pi} \int dE\, dE'\, dE_0\, 
P^*(E_0)\, P(E_0+E'-E)\,  \nonumber \\
&\quad \times e^{-i(E-E')t} \left[S^*(E)S(E') - 1\right] f_\mu(E-E_0).
\end{align}
To lowest order in the tunneling amplitude $s$, we expand
\begin{align}
S^*(E)S(E') - 1 &= -1 + e^{i \phi(E')}e^{-i \phi(E)} + 2i s\, e^{-i \phi(E)} \nonumber \\
& \times e^{i \phi(E')} \left[\sin\phi(E') - \sin\phi(E)\right] 
+ \mathcal{O}(s^2).
\end{align}
Thus, the first-order contribution to the current reads
\begin{align}
\left\langle I(t)\right\rangle^{(1)} 
=& -i\frac{e s}{\pi} \int dE\, dE'\, dE_0\, 
P^*(E_0)\, P(E_0+E'-E)\, \nonumber \\
& \times e^{-i(E-E')t}  e^{i \left[\phi(E')-\phi(E)\right]} f_\mu(E-E_0) \nonumber \\
& \times  \left[\sin\phi(E') - \sin\phi(E)\right].
\end{align}
We now shift the integration variable $E' \to E' + E$, such that $\phi(E'+E)-\phi(E)=E'd/u$, leading to
\begin{align}
\left\langle I(t)\right\rangle^{(1)} 
=& -i\frac{e s}{\pi} \int dE'\, dE_0\, dE\,
P^*(E_0)\, P(E_0+E')\,  \nonumber \\
& \times  \left[\sin\phi(E'+E) - \sin\phi(E)\right]  \nonumber \\
& \times e^{iE'\left(t - \frac{d}{u}\right)} f_\mu(E - E_0) .
\end{align}
Changing variables in the first term of the integrand $E'+E \to E$, we obtain
\begin{align}
\left\langle I(t)\right\rangle^{(1)} 
&= -i\frac{e s}{\pi} \int dE'\, dE_0\, 
P^*(E_0)\, P(E_0+E')\, \nonumber \\
& \times e^{iE'\left(t - \frac{d}{u}\right)} \int dE\, \sin\phi(E) \nonumber \\
& \times \left[f_\mu(E - E_0 - E') - f_\mu(E - E_0)\right].
\end{align}
Evaluating the energy integral yields
\begin{align}
&\int dE\, \sin\phi(E)\, \left[f_\mu(E - E_0 - E') - f_\mu(E - E_0)\right] 
\nonumber \\
&= \frac{2\pi}{\beta \sinh\left(\frac{\pi d}{u\beta}\right)} 
\sin\left(\mu \frac{d}{u} + E_0 \frac{d}{u} + \phi_0 + \frac{d}{u} \frac{E'}{2} \right)
\sin\left(\frac{d E'}{2u}\right).
\end{align}
Plugging this into the expression for the current and performing the shift $E'+E_0 \rightarrow E'$, we find
\begin{align}
&\left\langle I(t)\right\rangle^{(1)} 
= \frac{- 2i e s}{\beta \sinh\left(\frac{\pi d}{u\beta}\right)} 
\int dE'\, dE_0\, P^*(E_0)\, P(E')\,  \nonumber \\
&\quad \times e^{iE'\left(t - \frac{d}{u}\right)}\sin\left(\mu \frac{d}{u} + \phi_0 + \frac{d}{u} \frac{E'+E_0}{2} \right)
\sin\left(\frac{d E'}{2u}\right).
\end{align}
This result shows that, even within an alternative formalism, one recovers that at $\nu = 1$ the temperature-dependent prefactor factorizes out at lowest order in tunneling. The current can be rewritten as
\begin{align}
\left\langle I(t)\right\rangle^{(1)} 
=& \frac{i e s}{2\beta \sinh\left(\frac{\pi d}{u\beta}\right)} 
\sum_{\chi_1, \chi_2 = \pm} \chi_1 \chi_2\, 
e^{i \chi_1 \left(\phi_0 + \mu \frac{d}{u}\right)} \nonumber \\
& \times \int dE_0\, P^*(E_0)\, 
e^{-iE_0\left[\left(t - \frac{d}{u}\right) - \left(\chi_1+\chi_2\right) \frac{d}{2u}\right]} \nonumber \\
&\times
\int dE'\, P(E')\, 
e^{iE'\left[\left(t - \frac{d}{u}\right) - \left(\chi_1+\chi_2\right) \frac{d}{2u}\right]}.
\end{align}
Using the definition of $P(E)$ in Eq.~\eqref{eq:defP}, we finally express the current as
\begin{align}
\left\langle I(t)\right\rangle^{(1)} 
=& \frac{i e s}{2\beta \sinh\left(\frac{\pi d}{u\beta}\right)} 
\sum_{\chi_1, \chi_2 = \pm} \chi_1 \chi_2\, 
e^{i \chi_1 \left(\phi_0 + \mu \frac{d}{u}\right)} \nonumber \\
&\times \exp\left[
-i e \int_{\left(t - \frac{d}{u}\right) - (\chi_1 + \chi_2) \frac{d}{2u}}^{
\left(t - \frac{d}{u}\right) - \left(\chi_1+\chi_2\right) \frac{d}{2u}} d\tau\, V(\tau)
\right].
\end{align}
In the case of a delta-like pulse, the result simplifies to
\begin{align}
\left\langle I(t)\right\rangle^{(1)} 
&= \frac{-2i e s}{\beta \sinh\left(\frac{\pi d}{u\beta}\right)} 
\sum_{\chi_1, \chi_2 = \pm} \chi_1 \chi_2\, 
e^{i \chi_1 \left(\phi_0 + \mu \frac{d}{u}\right)} \nonumber \\
& \times e^{-2i\pi \left[ 
\text{sign}\left(\left(t - \frac{d}{u}\right) - \left(\chi_1 + \chi_2\right) \frac{d}{2u} \right) 
- \text{sign}\left(\left(t - \frac{d}{u}\right) - \left(\chi_1+\chi_2\right) \frac{d}{2u} \right)
\right]} \nonumber \\
&= 0,
\end{align}
as expected for $\nu = 1$, where no anyonic braiding effects contribute.

\section{First-Order Current as an Average Over Anyons}
\label{app:CurrentAnyons}
In this Appendix, we recast the first-order contribution to the charge current defined in Eq.~\eqref{eq:def_curr} as an average over anyons, in order to make the interpretation of the braiding mechanism more transparent. The current at first order reads:
\begin{align}
    \langle I(t)\rangle^{(1)} =& -i\frac{e u \sqrt{\nu}}{2\pi} \sum_{\eta} \eta \nonumber \\
    & \times \int dt_1 
    \left\langle T_K \left\{ \left[\partial_x \phi(x,t_-)\right]_{x=D} H_T(t_1^{\eta}) \right\} \right\rangle_{\Phi}.
\end{align}
Using Keldysh ordering, this can be recast as
\begin{align}
    \langle I(t)\rangle^{(1)} =& -i\frac{e u \sqrt{\nu}}{2\pi} \int dt_1\, 
    \Theta(t - t_1)  \nonumber \\
    & \times 
    \left\langle \left[\partial_x \phi(x,t), H_T(t_1) \right]_{x=D} \right\rangle_{\Phi}.
    \label{eq:def_currApp}
\end{align}
The commutator appearing in the equation above can be written as
\begin{align}
    \left[\partial_x \phi(x,t), H_T(t_1) \right] 
    &= -2\pi \sqrt{\nu} \sum_{\epsilon = \pm 1} e^{i\epsilon \kappa} \nonumber \\
        & \times \left[\Psi^{\dagger}\left(0, t_1-\epsilon \frac{d}{2u} \right) 
    \Psi\left( 0, t_1+\epsilon \frac{d}{2u} \right)\right] \nonumber \\
    & \times \big[ \delta \left(x - u t - \epsilon \frac{d}{2} + u t_1 \right) 
    \nonumber\\
    & - \delta \left(x - ut + \epsilon \frac{d}{2} + ut_1 \right) \big],
    \label{eq:CommTunneling}
\end{align}
where we used the commutation relations between the derivative of the bosonic field and the anyonic operators:
\begin{align}
    \left[ \partial_x \phi(x), \Psi^{\dagger}(x') \right] &= -2\pi \sqrt{\nu} \delta(x - x') \Psi^{\dagger}(x), \\
    \left[ \partial_x \phi(x), \Psi(x') \right] &= 2\pi \sqrt{\nu} \delta(x - x') \Psi(x).
\end{align}

Plugging Eq.~\eqref{eq:CommTunneling} into Eq.~\eqref{eq:def_currApp}, and converting the delta functions in space into delta functions in time, we obtain:
\begin{align}
    \langle I(t)\rangle^{(1)} &= i e^* \sum_{\epsilon = \pm 1} e^{i \epsilon \kappa} \nonumber \\
    & \times 
    \left\langle \left[
    \Psi^{\dagger}\left(D - ut, 0 \right) 
    \Psi\left(D - ut - \epsilon d, 0 \right)\right.\right. \nonumber \\
    &\left.\left.
    - \Psi^{\dagger}\left(D - ut + \epsilon d, 0 \right) 
    \Psi\left(D - ut, 0 \right)\right]
    \right\rangle_{\Phi}.
\end{align}

We now define the anyonic correlation function
\begin{equation}
    F_{\epsilon}(t) \equiv \left\langle 
    \left[\Psi^{\dagger}\left(0, t - \epsilon \frac{\tau_d}{2} \right)
    \Psi\left(0, t + \epsilon \frac{\tau_d}{2} \right) \right]
    \right\rangle_{\Phi},
\end{equation}
where we introduced $\tau_d = d/u$. Using this notation, the current becomes
\begin{equation}
    \langle I(t) \rangle^{(1)} 
    = i e^* \sum_{\epsilon = \pm 1} e^{i \epsilon \kappa} 
    F_{\epsilon}\left(t + \epsilon \frac{\tau_d}{2} - \frac{D}{u} \right) + \text{H.c.}
    \label{eq:def_curr_final}
\end{equation}
As a result, the current has been recast in terms of the anyon average
\begin{align}
F_{\pm} (t) =& \left\langle 0 \right| \Psi\left(0,0\right) \Psi^{\dagger}\left(0, t \mp \frac{\tau_d}{2} \right)\\&\times
    \Psi\left(0, t \pm \frac{\tau_d}{2} \right)  \Psi^{\dagger}\left(0,0\right)\left|0\right\rangle.
\end{align}

\section{Details about the calculation in the collider geometry}
\label{app:DetailsNoise}
In this Appendix, we compute the thermal average appearing in Eq.~\eqref{eq:ThermalAverageCollider} and the Poissonian average in Eq.~\eqref{eq:DefinitionBetaOmega}.
\subsection{Thermal average}
The thermal average that we aim at computing reads
\begin{align}
&\left\langle  T_K\left[I^{(0)}_1(t^-) \sum_{\eta=\pm}\eta~H^{(0)}_{T,\epsilon}(t_1^{\eta})\right] \right\rangle = \nonumber \\
&\qquad \frac{\Lambda e u\sqrt{\nu}}{4\pi^2 a}\Bigg\langle T_K\Bigg[\partial_x\phi^{(0)}_R(x,t^-)_{x=D_1} \nonumber \\
&\qquad \times\sum_{\eta=\pm}\eta e^{i\sqrt{\nu}\phi^{(0)}_R\left(\frac{\epsilon d}{2},t_1^{\eta}\right)}e^{-i\sqrt{\nu}\phi^{(0)}_R\left(\frac{-\epsilon d}{2},t_1^{\eta}\right)}\Bigg]\Bigg\rangle, 
\end{align}
where we replaced
\begin{equation}
H^{(0)}_{T,\epsilon}(t) = \frac{\Lambda}{2\pi a}e^{i\sqrt{\nu}\phi^{(0)}\left(\epsilon \frac{d}{2},t\right)}e^{-i \sqrt{\nu}\phi^{(0)}\left(-\epsilon\frac{d}{2},t\right)},
\end{equation}
from Eq.~\eqref{eq:BareTunnelingCollider} and
\begin{equation}
I^{(0)}_{1}(t) = \frac{e u \sqrt{\nu}}{2\pi}\left[\partial_x\phi^{(0)}_{R}(x,t)\right]_{x=D_{1/2}},
\end{equation}
from Eq.~\eqref{eq:BareCurrentCollider}. We recast $\partial_x\phi^{(0)}_{R}(x,t)$ as a derivative of a vertex operator 
\begin{align}
&\left\langle  T_K\left[I^{(0)}_1(t^-) \sum_{\eta=\pm}\eta~H^{(0)}_{T,\epsilon}(t_1^{\eta})\right] \right\rangle\nonumber=\\
&  =i\frac{\Lambda e u\sqrt{\nu}}{4\pi^2 a}\lim_{\gamma\rightarrow 0}\frac{1}{\gamma}\sum_{\eta=\pm}\eta\Bigg\langle T_K\Bigg[e^{-i\gamma \phi\left(x,t^-\right)}  e^{i\sqrt{\nu}\phi^{(0)}_R\left(\frac{\epsilon d}{2},t_1^{\eta}\right)}\nonumber \\
&\qquad \times e^{-i\sqrt{\nu}\phi^{(0)}_R\left(\frac{-\epsilon d}{2},t_1^{\eta}\right)}\Bigg]_{x=D_1}\Bigg\rangle. 
\end{align}
By means of the generalized Wick's theorem in Eq.~\eqref{eq:ProductBosonCorrelator}, we obtain
\begin{align}
& \left\langle  T_K\left[I^{(0)}_1(t^-) \sum_{\eta=\pm}\eta~H^{(0)}_{T,\epsilon}(t_1^{\eta})\right] \right\rangle = \frac{i\Lambda e u\sqrt{\nu}}{4\pi^2 a} \nonumber \\
& \quad \times \lim_{\gamma\rightarrow 0}\frac{1}{\gamma}\sum_{\eta=\pm}\eta e^{\delta G^{\eta \eta}\left(\epsilon d,0\right)}\partial_x e^{\gamma \sqrt{\nu}\epsilon \sum_{\xi}\xi G^{-\eta}\left(x,\xi \frac{d}{2};t,t_1\right)}. 
\end{align}
We compute the derivative and the limit, thus obtaining
\begin{align}
&\left\langle  T_K\left[I^{(0)}_1(t^-) \sum_{\eta=\pm}\eta~H^{(0)}_{T,\epsilon}(t_1^{\eta})\right] \right\rangle\nonumber=\\
& =i\frac{\Lambda e^* u\epsilon}{4\pi^2 a}e^{\delta \mathcal{G}\left(\frac{d}{u}\right)} \sum_{\substack{\eta=\pm \\ \xi=\pm}}\eta\xi \left. \partial_x G^{-\eta}\left(x,\xi \frac{d}{2};t,t_1\right)\right|_{x=D_1},
\end{align}
where we used the property in Eq.~\eqref{eq:UsefulRelation1b}. We use the formula for the derivative of the boson Green's function in Eq.~\eqref{eq:UsefulRelation3}
\begin{align}
&\left\langle  T_K\left[I^{(0)}_1(t^-) \sum_{\eta=\pm}\eta~H^{(0)}_{T,\epsilon}(t_1^{\eta})\right] \right\rangle\nonumber=\\
&=i\frac{\Lambda e^* u\epsilon}{4\pi^2 a}e^{\delta \mathcal{G}\left(\frac{d}{u}\right)}\sum_{\substack{\eta=\pm \\ \xi=\pm}}\eta\xi \left\{\partial_x \mathcal{G}\left(t-t_1-\frac{x-\xi \frac{d}{2}}{u}\right)\right. \nonumber \\
& \qquad \left.+\frac{i\pi}{u}\left[\eta \Theta(x)+\Theta(-x) \right]\delta\left(t-t_1-\frac{x-\xi \frac{d}{2}}{u}\right)\right\}_{x=D_1} =\nonumber \\
&-\frac{e^*\Lambda  e^{\delta \mathcal{G}\left(\frac{d}{u}\right)}}{2\pi a}\epsilon\sum_{\xi=\pm}\xi \delta\left(t-t_1-\frac{D_1-\xi \frac{d}{2}}{u}\right),
\end{align}
where in the last step we used the fact that $x=D_1>0$ and that $\sum_{\eta}\eta = 0$. The above equation corresponds to the result presented in Eq.~\eqref{eq:ThermalAverageCollider} of the main text.

\subsection{Poissonian average}
Here, we compute the quantity defined in Eq.~\eqref{eq:DefinitionBetaOmega}, which corresponds to the Fourier transform of the expression
\begin{align}
\bar{B}\left(t_1-t_2\right)&\equiv\sum_{\epsilon=\pm}e^{i\epsilon\kappa}\epsilon \left\langle \mathcal{B}_{\epsilon}(t_1)\,\partial_{t_2} \delta N \left(t_2\right) \right\rangle_P=\nonumber\\& 2i \Im\left[e^{i\kappa}\left\langle \mathcal{B}_{+}(t_1)\,\partial_{t_2} \delta N \left(t_2\right) \right\rangle_P \right] ,\label{eq:DefinitionBetaBar}
\end{align}
where the operator $\mathcal{B}_{\epsilon}(t)$, defined in Eq.~\eqref{eq:DefinitionBraidingFactor}, reads
\begin{equation}
\mathcal{B}_{\epsilon}(t) = \exp\left\{i 2\pi \epsilon \lambda \sum_{\chi=\pm}\chi\,\delta N\left(t-\frac{\chi d}{2u}\right)\right\}.
\end{equation}
Our goal is to evaluate the Poissonian average in Eq.~\eqref{eq:DefinitionBetaBar}. We are thus led to compute the following general quantity
\begin{equation}
\left\langle \partial_{\tau_0}\delta N(\tau_0)\exp\left\{i2\pi\lambda \left[\delta N(\tau_1)-\delta N(\tau_2)\right]\right\} \right\rangle{_P},
\label{eq:PoissonAverageCollider}
\end{equation}
where the times $\tau_1$ and $\tau_2$ correspond to the arguments of the anyon numbers in $\mathcal{B}_+(t_1)$. The evaluation of this expression requires care in separating the time {intervals} into non-overlapping regions, since the average is performed over a Poissonian distribution of uncorrelated anyon emission events. {This separation} depends on the relative ordering of the times $\tau_0$, $\tau_1$ and $\tau_2$.

When $\tau_0$ lies outside the interval $[\tau_1, \tau_2]$, either to the left or to the right, {one can separate the contribution in $\tau_0$ from the rest}. In such situations, the expectation value factorizes, and we may evaluate each term separately.

For instance, when $\tau_0 < \tau_1 < \tau_2$, one has:
\begin{align}
&\left\langle \partial_{\tau_0}\delta N(\tau_0)\exp\left\{i2\pi\lambda \left[\delta N(\tau_1)-\delta N(\tau_2)\right]\right\} \right\rangle{_P} \nonumber \\
& \quad = \left\langle \partial_{\tau_0}\delta N(\tau_0)\right\rangle\left\langle\exp\left\{i2\pi\lambda \left[\delta N(\tau_1)-\delta N(\tau_2)\right]\right\} \right\rangle{_P}.
\end{align}
By using the relation $\left\langle \partial_{\tau_0}\delta N(\tau_0) \right\rangle{_P} = \gamma$ for the average Poissonian emission rate {along with} the average in Eq.~\eqref{eq:Poisson}, one finds for $\tau_0 \notin \left[\tau_1,\tau_2\right]$
\begin{align}
& \left\langle \partial_{\tau_0}\delta N(\tau_0)\exp\left\{i2\pi\lambda \left[\delta N(\tau_1)-\delta N(\tau_2)\right]\right\} \right\rangle{_P} \nonumber \\
& \qquad = \gamma\, e^{-\gamma(\tau_2 - \tau_1)(1 - e^{-i2\pi\lambda})}.
\end{align}

When instead the time $\tau_0$ lies between $\tau_1$ and $\tau_2$, more care is required to split the average into independent parts. One can recast the expression in Eq.~\eqref{eq:PoissonAverageCollider} as
\begin{align}
&\left\langle \partial_{\tau_0}\delta N(\tau_0)\exp\left\{i2\pi\lambda \left[\delta N(\tau_1)-\delta N(\tau_2)\right]\right\} \right\rangle{_P} \nonumber \\
& \quad = \partial_{\tau_0}\left\langle \left[\delta N(\tau_0) - \delta N(\tau_1)\right] e^{i2\pi\lambda \left[\delta N(\tau_1)-\delta N(\tau_0)\right]}\right.\nonumber \\
& \qquad \left.\times e^{i2\pi\lambda \left[\delta N(\tau_0)-\delta N(\tau_2)\right]} \right\rangle{_P}  .
\end{align}

By noticing that the intervals $\left[\tau_1,\tau_0\right)$ and $\left(\tau_0,\tau_2\right]$ are not overlapping, the average can be consequently split as
\begin{align}
& \partial_{\tau_0}\left\langle \left[\delta N(\tau_0) - \delta N(\tau_1)\right] e^{i2\pi\lambda \left[\delta N(\tau_1)-\delta N(\tau_0)\right]}\right\rangle{_P} \nonumber \\
& \qquad \times \left\langle e^{i2\pi\lambda \left[\delta N(\tau_0)-\delta N(\tau_2)\right]} \right\rangle{_P}  .\label{eq:AverageInternalTimes}
\end{align}
Then, we notice that
\begin{align}
&\left\langle \left[\delta N(\tau_0) - \delta N(\tau_1)\right] \exp\left\{i2\pi\lambda \left[\delta N(\tau_1)-\delta N(\tau_0)\right]\right\}\right\rangle{_P} =\nonumber\\
& \qquad = \frac{1}{2\pi i}\partial_{\lambda} \left\langle \exp\left\{i2\pi\lambda \left[\delta N(\tau_1)-\delta N(\tau_0)\right]\right\}\right\rangle{_P}.
\end{align}
By exploiting this identity in Eq.~\eqref{eq:AverageInternalTimes}, one finds for $\tau_0 \in \left[\tau_1,\tau_2\right]$
\begin{align}
&\left\langle \partial_{\tau_0}\delta N(\tau_0)\exp\left\{i2\pi\lambda \left[\delta N(\tau_1)-\delta N(\tau_2)\right]\right\} \right\rangle{_P} \nonumber \\
& \qquad = \gamma\, \, e^{-i2\pi\lambda}e^{-\gamma(\tau_2 - \tau_1)(1 - e^{-i2\pi\lambda})}.
\end{align}

Finally, by specializing the time arguments to those appearing in Eq.~\eqref{eq:DefinitionBetaBar}, namely $\tau_0 = t_2$, $\tau_1 = t_1 - \tau_d/2$, and $\tau_2 = t_1 + \tau_d/2$, we obtain the expression:
\begin{align}
&\left\langle \mathcal{B}_+(t_1)\partial_{t_2}\delta N(t_2)\right\rangle{_P} = \gamma\, \, e^{-\gamma \tau_d (1 - e^{-i2\pi\lambda})}\nonumber\\
& {\times} \left[1-\left(1-e^{-i2\pi\lambda}\right)\Theta\left(t_1+\frac{\tau_d}{2}-t_2\right)\Theta\left(t_2+\frac{\tau_d}{2}-t_1\right)\right].
\end{align}
This expression encodes the full time-dependent behavior of the correlator $\bar{B}(t_1 - t_2)$. By plugging this expression back into Eq.~\eqref{eq:DefinitionBetaBar}, we obtain
\begin{align}
&\bar{B}\left(t_1 - t_2\right) = -2 i \gamma
\, e^{-\gamma \tau_d \left[1 - \cos\left(2\pi \lambda\right)\right]}  \nonumber \\
& \quad \times
 \Theta\left(t_1 - t_2 + \frac{\tau_d}{2}\right) 
\, \Theta\left(t_2 - t_1 + \frac{\tau_d}{2}\right) \nonumber \\
&\quad \times \Bigg\{ 
\cos\left[\kappa - \gamma \tau_d \sin\left(2\pi\lambda\right)\right] 
\sin\left(2\pi \lambda\right)  \nonumber \\
& \qquad+ \sin\left[\kappa - \gamma \tau_d \sin\left(2\pi\lambda\right)\right] 
\left[1 -\cos\left(2\pi \lambda\right)\right] 
\Bigg\} \nonumber \\
&\quad + 2 i \gamma 
\, e^{-\gamma \tau_d \left[1 - \cos\left(2\pi \lambda\right)\right]} 
\, \sin\left[\kappa - \gamma \tau_d \sin\left(2\pi\lambda\right)\right] .
\label{eq:Bbar2}
\end{align}
To compute the Fourier transform of $\bar{B}(t)$, we note that
\begin{align}
&\int d\left(t-t'\right)e^{i \omega \left(t-t'\right)}\Theta\left(t - t' + \frac{\tau_d}{2}\right) 
\, \Theta\left(t' - t + \frac{\tau_d}{2}\right)\nonumber\\& = \frac{2}{\omega}\sin\left(\omega \frac{\tau_d}{2}\right),
\end{align}
and
\begin{align}
&\int d\left(t-t'\right)e^{i \omega \left(t-t'\right)}= \delta(\omega).
\end{align}
By inserting Eq.~\eqref{eq:Bbar2} into Eq.~\eqref{eq:DefinitionBetaOmega} and using these two results, we arrive at
\begin{align}
\mathcal{B}\left(\omega\right)
&= -2i\gamma \tau_d e^{-\gamma \tau_d\left[1-\cos\left({2\pi \lambda}\right)\right]} \frac{ \sin \left( \omega \frac{\tau_d}{2} \right)}{ \omega \frac{\tau_d}{2}} \nonumber \\
& \times \left\{ \cos\left[\kappa - \gamma\tau_d\sin\left(2\pi\lambda\right)\right]\sin\left(2\pi \lambda\right)\right.\nonumber\\&\left. +\sin\left[\kappa - \gamma \tau_d\sin\left(2\pi\lambda\right)\right]\left[1-\cos\left(2\pi \lambda\right)\right] \right\}
\nonumber \\
& + 2 i \gamma e^{-\gamma \tau_d\left[1-\cos\left({2\pi \lambda}\right)\right]}  \sin\left[\kappa- \gamma \tau_d\sin\left(2\pi\lambda\right)\right] \delta (\omega).
\label{eq:FT}
\end{align}
The former expression can be recast as
\begin{align}
&\mathcal{B}\left(\omega\right) = -4i  \frac{ \sin \left( \omega \frac{\tau_d}{2} \right)}{ \omega \frac{\tau_d}{2}} \mathcal{C}_\lambda (\tilde{\gamma})\nonumber\\& + 2 i  \gamma e^{-\tilde{\gamma}\left[1-\cos\left({2\pi \lambda}\right)\right]}  \sin\left[\kappa- \tilde{\gamma}\sin\left(2\pi\lambda\right)\right] \delta (\omega) ,
\label{eq:FT2}
\end{align}
where we introduced
\begin{align}
    \tilde{\gamma} &\equiv \gamma \tau_d ,\\
    \mathcal{C}_\lambda (\tilde{\gamma}) &\equiv \sin \left(\pi  \lambda \right) \tilde{\gamma} e^{-\tilde{\gamma} \left[1-\cos\left({2\pi \lambda}\right)\right]}\cos \left[\tilde{\gamma} \sin\left(2\pi\lambda\right)+\pi\lambda - \kappa\right],
\end{align}
and where we used the trigonometric relations
\begin{align}
&\sin\left(2\theta\right) = 2 \sin\theta\cos\theta,\\
&1-\cos\left(2\theta\right) = 2 \sin^2\theta,\\
&\cos\theta_1\cos\theta_2+\sin\theta_1\sin\theta_2 = \cos\left(\theta_2-\theta_1\right).
\end{align}
\bibliographystyle{apsrev4-2}
\bibliography{AnyonsBibliography}

\end{document}